\documentclass[aps,showpacs,preprintnumbers,amsmath, amssymb]{revtex4}

\usepackage{float}
\usepackage{graphics,epsfig}
\usepackage{graphicx}
\usepackage{dcolumn}
\usepackage{bm}

\begin{document}
\baselineskip=0.8 cm
\title{{\bf Holographic insulator/superconductor phase transitions with excited states}}

\author{Liang OuYang$^{1}$, Dong Wang$^{1}$, Xiongying Qiao$^{1}$, Mengjie Wang$^{1}$\footnote{mjwang@hunnu.edu.cn}, Qiyuan Pan$^{1,2}$\footnote{panqiyuan@hunnu.edu.cn}, and Jiliang Jing$^{1,2}$\footnote{jljing@hunnu.edu.cn}}
\affiliation{$^{1}$Key Laboratory of Low Dimensional Quantum Structures and Quantum Control of Ministry of Education, Synergetic Innovation Center for Quantum Effects and Applications, and Department of Physics, Hunan Normal University, Changsha, Hunan
410081, China} \affiliation{$^{2}$Center for Gravitation and Cosmology, College of Physical Science and Technology, Yangzhou University, Yangzhou 225009, China}

\vspace*{0.2cm}
\begin{abstract}
\baselineskip=0.6 cm
\begin{center}
{\bf Abstract}
\end{center}

We construct a family of solutions of the holographic insulator/superconductor phase transitions with the excited states in the AdS soliton background by using both the numerical and analytical methods. The interesting point is that the improved Sturm-Liouville method can not only analytically investigate the properties of the phase transition with the excited states, but also the distributions of the condensed fields in the vicinity of the critical point. We observe that, regardless of the type of the holographic model, the excited state has a higher critical chemical potential than the corresponding ground state, and the difference of the dimensionless critical chemical potential between the consecutive states is around 2.4, which is different from the finding of the metal/superconductor phase transition in the AdS black hole background. Furthermore, near the critical point, we find that the phase transition of the systems is of the second order and a linear relationship exists between the charge density and chemical potential for all the excited states in both s-wave and p-wave insulator/superconductor models.

\end{abstract}


\pacs{11.25.Tq, 04.70.Bw, 74.20.-z}

\maketitle
\newpage
\vspace*{0.2cm}

\section{Introduction}

The gauge/gravity correspondence, which was first proposed by Maldacena \cite{Maldacena}, relates the large $N$ limit of
the strongly coupled gauge theories to the classical gravitational theories in the anti-de Sitter (AdS) spacetime. As a version of this duality, the anti-de Sitter/conformal field theories (AdS/CFT) correspondence \cite{Witten,Gubser1998} has been used to study a variety of the condensed matter systems over the past ten years \cite{JZaanen,LandsteinerYY}, especially for the high-temperature superconductor systems \cite{HartnollJHEP12}. With this holographic duality, Gubser suggested that an instability may be trigged by a scalar field around a charged AdS black hole, which is dual to a superconducting phase transition \cite{GubserPRD78}. Later on, Hartnoll \emph{et al.} built the first holographic superconductor model to reproduce the properties of a ($2+1$)-dimensional s-wave superconductor \cite{HartnollPRL101}. Due to the potential applications to the condensed matter
physics, a large number of the s-, p- and d-wave holographic superconductor models in different theories of
gravity have been constructed; see reviews~\cite{HartnollRev,HerzogRev,HorowitzRev,CaiRev} and references therein.

In most cases, the studies on the holographic superconductors focus on the ground state, which is the most stable mode. It is of great interest to investigate the excited states of the superconductors by holography because of the potential significance in the condensed matter physics \cite{BCS,Peeters2000,Vodolazov2002,RMP2004,RMP2011,LiuSonner}. Analyzing the colorful horizons with the charge in the AdS space, Gubser showed the existence of branches of solutions with multiple nodes corresponding to the excited states \cite{GubserPRL2008}. By constructing numerical solutions of the holographic s-wave superconductors with the excited states in the probe limit \cite{WangJHEP2020} and away from the probe limit \cite{WangLLZ}, Wang \emph{et al.} found that the excited state has a lower critical temperature than the corresponding ground state, which has recently been confirmed analytically by us, through using the improved variational method for the Sturm-Liouville eigenvalue problem \cite{QiaoEHS}. Li \emph{et al.} discussed the non-equilibrium condensation process of the holographic s-wave superconductor with the excited states as the intermediate states during the relaxation, and observed the dynamical formation of the excited states as the intermediate states during the relaxation from the normal state to the ground state, which may provide the useful information about the non-equilibrium superconductivity that can be compared to the experiments in the condensed matter physics \cite{LiWWZ}. More recently, Xiang \emph{et al.} generalized the study of the holographic superconductors with the excited states to the framework of massive gravity and obtained the effects of the graviton mass on the scalar condensate and conductivity of the system \cite{XiangZW}.

The aforementioned works on the holographic superconductors with the excited states were implemented on the backgrounds of AdS black hole. Thus, it is interesting to study whether there exist the solutions of the holographic dual models with the excited states in the backgrounds of AdS soliton. For the ground state, Nishioka \emph{et al.} showed that the soliton becomes unstable to form scalar hair and a second order phase transition can happen when the chemical potential is sufficiently large beyond a critical value $\mu_{c}$, which can be used to describe the transition between the insulator and superconductor \cite{Nishioka-Ryu-Takayanagi}. This pioneering work has led to many investigations concerning the s-wave \cite{Pan-Wang,HorowitzSolition,BrihayeSolition,CaiLZSolition,PengPWSolition,CaiLZZ2011Solition,MontullSolition,LeeSolition,
CaiHLZSolition,KuangLWSolition,CaiHLLSolition,BasuDDNSolition,ErdmengerGPSolition,QiBXGSolition,PengPLSolition,PengLPLBSolition,
ParaiGGSolition,LuWMLDSolition,ParaiGGEPJCSolition,LuLWSolition}, p-wave \cite{AkhavanPWaveSolition,Pan2011PWaveSolition,RoychowdhuryPWaveSolition,CaiHLLRPWaveSolition,CaiHLLWPWaveSolition,
RogatkoWPWaveSolition,LaiPJW2016,LuWDLPWaveSolition,LvPLB2020,LaiHPJPWaveSolition} and s+p \cite{LiZZSolition} holographic insulator/superconductor phase transition with the ground state. Interestingly, taking advantage of the Sturm-Liouville method first proposed by Siopsis and Therrien \cite{Siopsis}, Li studied the holographic insulator/superconductor phase transition analytically and obtained not only the ground state of the phase transition but also the first excited state in Ref. \cite{HFLi}. As a further step along this line, in this work we first construct the novel solutions of the holographic insulator/superconductor phase transitions with the excited states by using the numerical shooting method, and then, following our previous work \cite{QiaoEHS}, employ the Sturm-Liouville method by including more higher order terms in the expansion of the trial function, to study the higher excited state and back up the numerical computations. Considering that the probe limit can simplify the problem but retain most of the interesting physics since the nonlinear interactions between the scalar (or vector) and Maxwell field are retained, we will concentrate on this probe limit and consider the five-dimensional Schwarzschild-AdS soliton background in the form
\begin{eqnarray}\label{soliton}
ds^2=-r^{2}dt^{2}+\frac{dr^2}{f(r)}+f(r)d\varphi^2+r^{2}(dx^{2}+dy^{2}),
\end{eqnarray}
where $f(r)=r^{2}(1-r_{s}^{4} /r^{4})$ with a conical singularity $r_{s}$, i.e., the tip of the soliton. It should be noted that one can impose a period $\beta_{s}=\pi/r_{s}$ for the coordinate $\varphi$ to remove the singularity.

This paper is organized as follows. In Sec. II we explore the solutions of the s-wave holographic insulator/superconductor phase transition with the excited states by using the shooting method and the generalized Sturm-Liouville method. In Sec. III, we study the solutions of the p-wave case via the Maxwell complex vector field model \cite{CaiPWave-1,CaiPWave-2}, by using the aforementioned numerical and analytical methods. We conclude in the last section with our main results.

\section{Excited states of the s-wave holographic insulator/superconductor phase transition}

In order to study the solutions of the s-wave holographic dual model with the excited states in the AdS soliton background, we begin with a Maxwell field coupled to a charged complex scalar field via the action
\begin{eqnarray}
S=\int d^{5}x\sqrt{-g}\left[-\frac{1}{4}F_{\mu\nu}F^{\mu\nu}-g^{\mu\nu}(\nabla_{\mu}\psi-iqA_{\mu}\psi)(\nabla_{\nu}\psi-iqA_{\nu}\psi)^{\ast}
-m^{2}|\psi|^{2} \right],
\end{eqnarray}
where $q$ and $m$ are the charge and mass of the scalar field $\psi$, respectively. By adopting the ansatz for the matter fields $\psi=\psi(r)$ and $A_{t}=\phi(r)$, we get the equations of motion
\begin{eqnarray}
\psi^{\prime\prime}+\left(\frac{3}{r}+
\frac{f^\prime}{f}\right)\psi^\prime
-\frac{1}{f}\left(m^2-\frac{q^{2}\phi^2}{r^2}\right)\psi=0\;, \label{psi}
\end{eqnarray}
\begin{eqnarray}
\phi^{\prime\prime}+\left(\frac{1}{r}+\frac{f^\prime}{f}\right)
\phi^\prime-\frac{2q^{2}\psi^2}{f}\phi=0\;, \label{phi}
\end{eqnarray}
where the prime denotes the derivative with respect to $r$. In order to get the solutions in the superconducting phase, we impose the boundary conditions at the tip $r_{s}$ by requiring the matter fields to be regular. In the asymptotic AdS region $r\rightarrow\infty$, the solutions behave as
\begin{eqnarray}
\psi=\frac{\psi_{-}}{r^{\Delta_{-}}}+\frac{\psi_{+}}{r^{\Delta_{+}}}\,,\hspace{0.5cm}
\phi=\mu-\frac{\rho}{r^{2}}\,, \label{infinity}
\end{eqnarray}
with the characteristic exponents $\Delta_\pm=2\pm\sqrt{4+m^{2}}$. Here, $\mu$ and $\rho $ are interpreted as the chemical potential and charge density in the dual field theory, respectively. From the AdS/CFT correspondence, the coefficients $\psi_{-}$ and $\psi_{+}$ both multiply normalizable modes of the scalar field equations and correspond to the vacuum expectation values
$<\mathcal{O}_{-}>=\psi_{-}$, $<\mathcal{O}_{+}>=\psi_{+}$ of an operator $\mathcal{O}$ dual to the scalar field. We can impose boundary condition that either $\psi_{+}$ or $\psi_{-}$ vanishes, just as in \cite{HartnollJHEP12,HartnollPRL101}. Since the choices of the scalar field mass do not modify results qualitatively, in this section we set $m^{2}=-15/4$ for concreteness.

\subsection{Numerical analysis}

We first use the shooting method to solve Eqs. (\ref{psi}) and (\ref{phi}) numerically. It should be noted that the following scaling transformations
\begin{eqnarray}\label{PWSSymmetry}
&&r\rightarrow\lambda r,~~(t, \varphi, x,
y)\rightarrow\frac{1}{\lambda}(t, \varphi, x,
y),~~(q,\psi)\rightarrow (q,\psi),~~\phi\rightarrow\lambda\phi,\nonumber \\
&&\mu\rightarrow\lambda\mu,~~
\rho\rightarrow\lambda^{3}\rho,~~\psi_{\pm}\rightarrow\lambda^{\Delta_{\pm}}\psi_{\pm},
\end{eqnarray}
where $\lambda$ is a positive number, are held for Eqs.~\eqref{psi} and~\eqref{phi}. Therefore, we can set $r_{s}=1$, $q=1$ and introduce a dimensionless quantity $z=r_{s}/r$ when performing numerical calculations.

In Ref. \cite{Nishioka-Ryu-Takayanagi}, Nishioka~\emph{et al.} pointed out that, in the ground
state, there is a critical chemical potential $\mu_{c}$, above which the solution is unstable and a hair can be developed; while for $\mu<\mu_{c}$ the scalar field is zero and it can be interpreted as the insulator phase. This means that there is a phase transition between the insulator and superconductor phases around the critical chemical potential $\mu_{c}$. Following the same spirit, here we numerically solve Eqs.~\eqref{psi} and~\eqref{phi}, to explore excited states of the s-wave holographic insulator/superconductor phase transition.

\begin{figure}[H]
\includegraphics[scale=0.65]{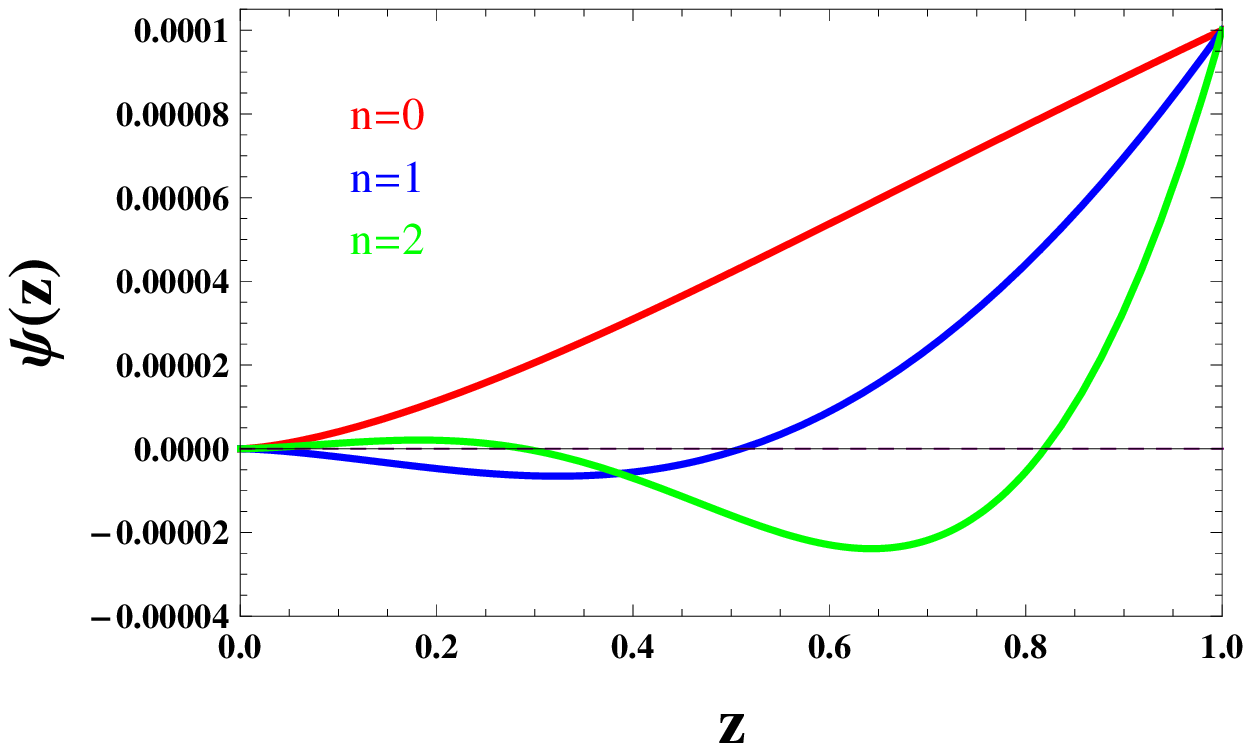}\hspace{0.2cm}%
\includegraphics[scale=0.65]{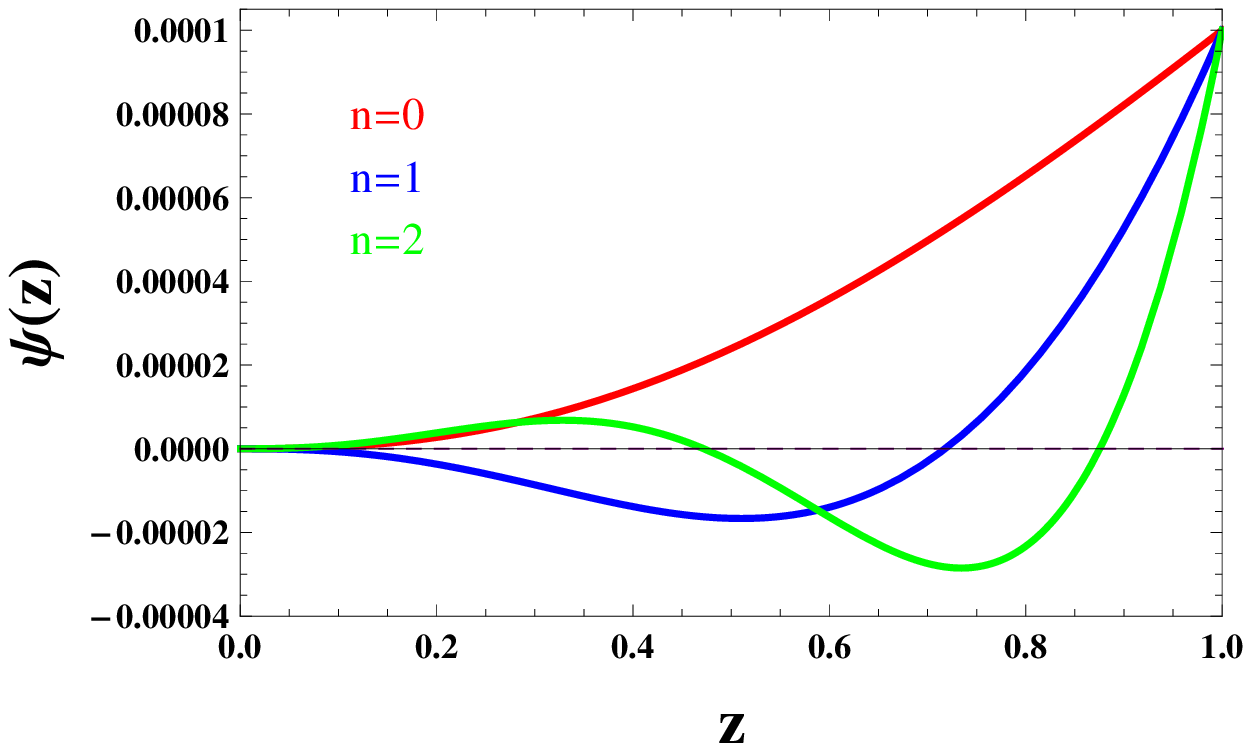}\\ \vspace{0.0cm}
\caption{\label{Psiz} (Color online) The scalar field $\psi(z)$ as a function of the radial coordinate $z$ outside the horizon with the scalar operators $\mathcal{O}_{-}$ (left) and $\mathcal{O}_{+}$ (right) for the fixed mass of the scalar field $m^{2}=-15/4$ by using the numerical shooting method. In each panel, the three lines from top to bottom correspond to the ground ($n=0$, red), first ($n=1$, blue) and second ($n=2$, green) excited states, respectively.}
\end{figure}

In Fig. \ref{Psiz}, we use the numerical shooting method to plot the distribution of the scalar field $\psi(z)$ as a function of $z$ for the scalar operators $\mathcal{O}_{-}$ and $\mathcal{O}_{+}$ with the fixed scalar field mass $m^{2}=-15/4$. In each panel, the red line has no intersecting points with the $\psi(z)=0$ axis at nonvanishing $z$, which denotes the ground state with the number of nodes $n=0$. And the blue line has one intersecting point with $\psi(z)=0$ axis while the green line has two, corresponding to the first ($n=1$) and second ($n=2$) states respectively, which shows that the $n$-th excited state has exactly $n$ nodes.

\begin{table}[ht]
\caption{\label{SWave} The critical chemical potential $\mu_{c}$ obtained by the shooting method for the scalar operators $\mathcal{O}_{-}$ and $\mathcal{O}_{+}$ with the fixed mass of the scalar field $m^{2}=-15/4$ from the ground state to the sixth excited state.}
\begin{tabular}{c c c c c c c c}
         \hline
$n$ & 0 & 1 & 2 & 3 & 4 & 5 & 6
        \\
        \hline
~~~~$\mathcal{O}_{-}$~~~~&~~~~~$0.836$~~~~~&~~~~~$3.055$~~~~~&~~~~~$5.427$~~~~~&~~~~~$7.816$~~~~
&~~~~~$10.209$~~~~&~~~~~$12.604$~~~~&~~~~~$15.000$~~~~
          \\
~~~~$\mathcal{O}_{+}$~~~~&~~~~~$1.888$~~~~~&~~~~~$4.234$~~~~~&~~~~~$6.616$~~~~~&~~~~~$9.005$~~~~
&~~~~~$11.397$~~~~&~~~~~$13.791$~~~~&~~~~~$16.186$~~~~
          \\
        \hline
\end{tabular}
\end{table}

In order to obtain the effect of the node number on the critical chemical potential for the scalar operators $\mathcal{O}_{-}$ and $\mathcal{O}_{+}$, in Table \ref{SWave} we give the critical chemical potential $\mu_{c}$ obtained by the shooting method with the fixed mass of the scalar field $m^{2}=-15/4$ from the ground state to the sixth excited state. We observe that the critical chemical potential $\mu_{c}$ increases with increasing the number of nodes $n$ for both the operators $\mathcal{O}_{-}$ and $\mathcal{O}_{+}$, i.e., an excited state has a higher critical chemical potential than the ground state, which indicates that the ground state is the first one to condense when increasing the chemical potential. Using the numerical results obtained by the shooting method, we can express the relation between $\mu_{c}$ and $n$ as
\begin{eqnarray}\label{SWaveMuc}
\mu_{c}\approx
\left\{
\begin{array}{rl}
2.370n+0.739, &  \quad {\rm for} \ \mathcal{O}_{-}\,,\\ \\
2.385n+1.861, &  \quad {\rm for} \  \mathcal{O}_{+}\,,
\end{array}\right.
\end{eqnarray}
which shows that, for both operators, the dimensionless critical chemical potential becomes evenly spaced for the number of nodes $n$ and the difference of the dimensionless critical chemical potential $\mu_{c}$ between the consecutive states is around 2.4.

\begin{figure}[H]
\includegraphics[scale=0.65]{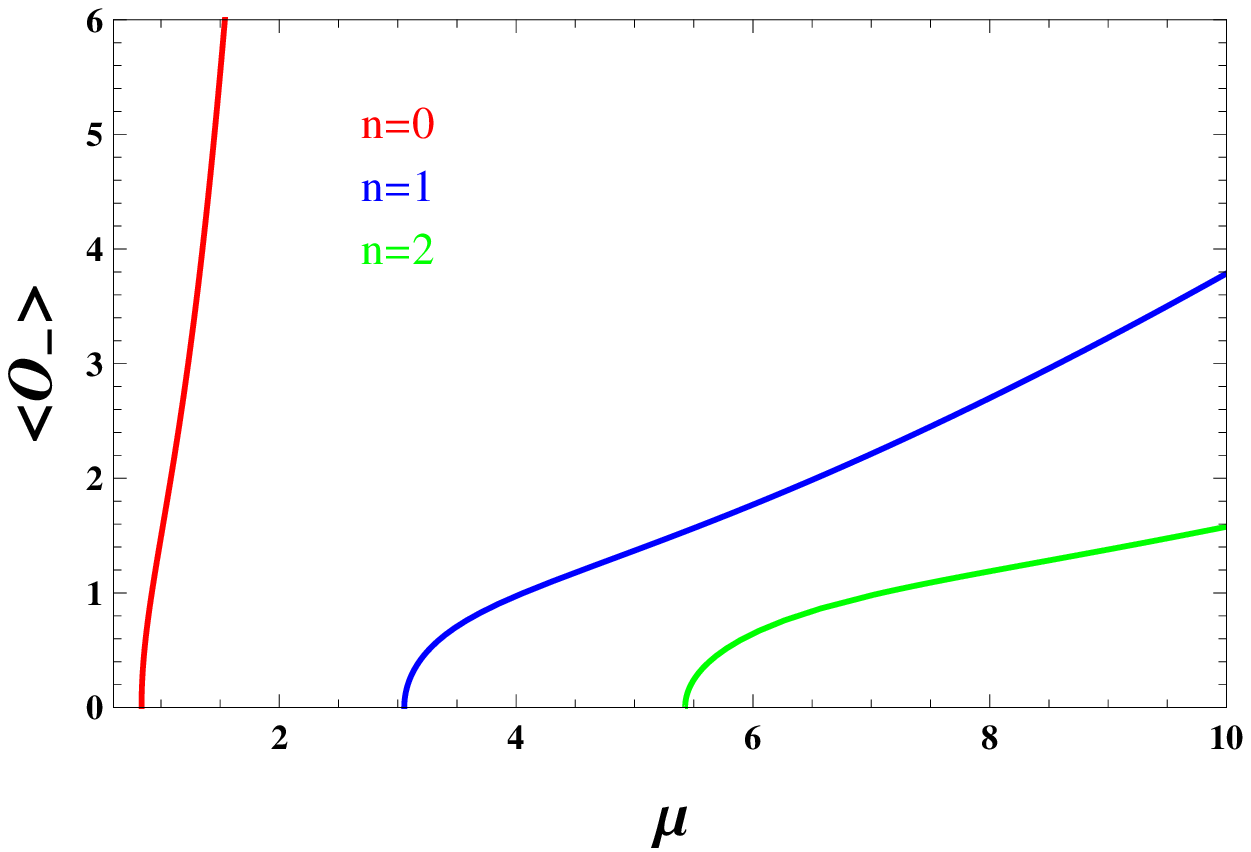}\hspace{0.2cm}%
\includegraphics[scale=0.65]{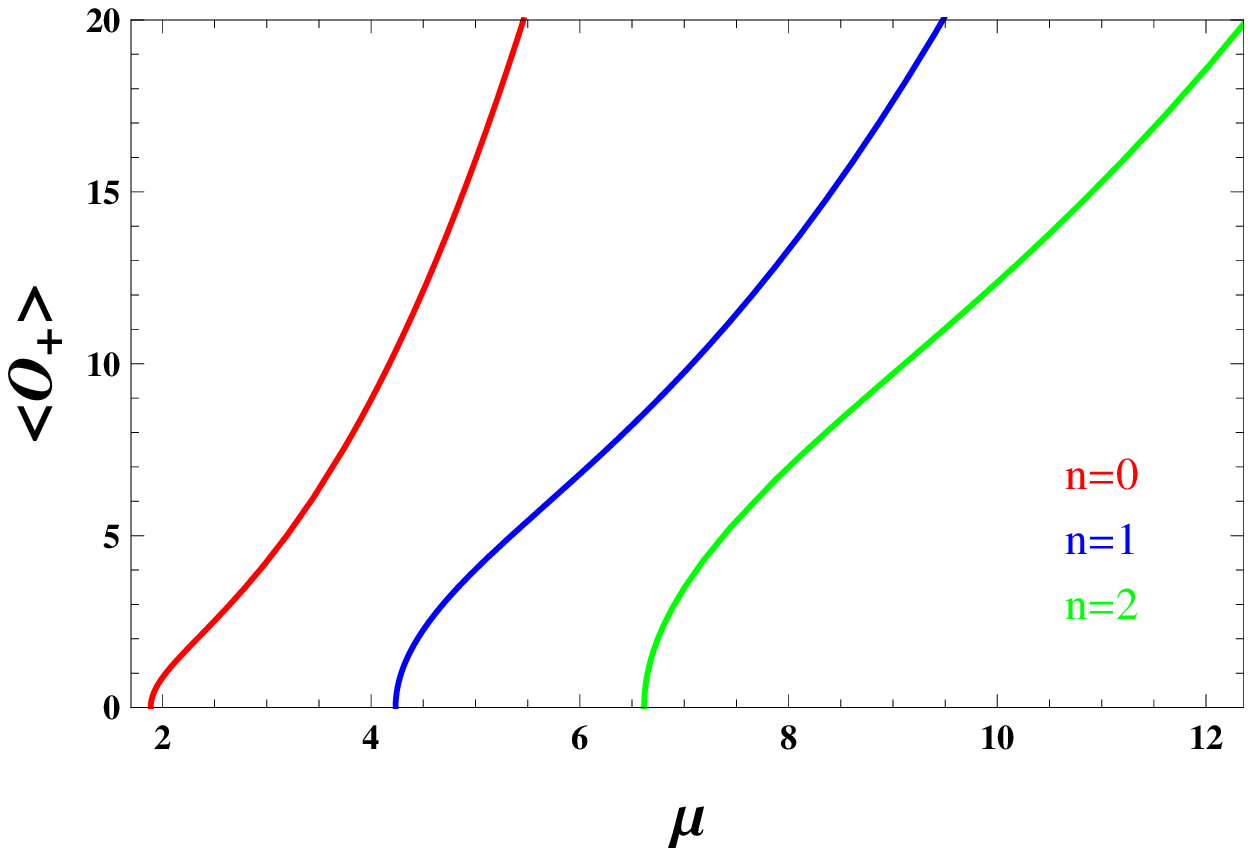}\\ \vspace{0.0cm}
\caption{\label{CondSWave} (Color online) The condensates of
the scalar operators $\mathcal{O}_{+}$ and $\mathcal{O}_{-}$ with respect to the chemical potential $\mu$ for the mass of the scalar field $m^{2}=-15/4$. In each panel, the three lines from left to right correspond to the ground ($n=0$, red), first ($n=1$, blue) and second ($n=2$, green) excited states, respectively.}
\end{figure}

In Fig. \ref{CondSWave}, we present the condensates of the scalar operators $\mathcal{O}_{-}$ and $\mathcal{O}_{+}$ as a function of the chemical potential with the mass of the scalar field $m^{2}=-15/4$ for the first three lowest-lying modes $n=0$, $1$ and $2$. For both the scalar operators $\mathcal{O}_{-}$ and $\mathcal{O}_{+}$, the behavior of the condensates for the excited states is similar to that for the ground state in the probe limit \cite{Nishioka-Ryu-Takayanagi}. Moreover, we observe that, similar to the ground state, a phase transition can happen when the chemical potential is over
a critical value $\mu_{c}$ for an excited state, which can be used to describe the transition between the
insulator and superconductor with the excited state. By fitting these curves for small condensate, we obtain
\begin{eqnarray}\label{SWaveOFu}
\langle{\cal O}_{-}\rangle\approx \left\{
                                      \begin{array}{ll}
                                        2.866\left(\frac{\mu}{\mu^{(0)}_{c}}-1\right)^{1/2}, & \hbox{Ground state,} \\
                                        2.030\left(\frac{\mu}{\mu^{(1)}_{c}}-1\right)^{1/2}, & \hbox{1st excited state,} \\
                                        2.015\left(\frac{\mu}{\mu^{(2)}_{c}}-1\right)^{1/2}, & \hbox{2st excited state,}
                                      \end{array}
                                    \right.
\end{eqnarray}
and
\begin{eqnarray}\label{SWaveOZheng}
\langle{\cal O}_{+}\rangle\approx \left\{
                                      \begin{array}{ll}
                                        3.400\left(\frac{\mu}{\mu^{(0)}_{c}}-1\right)^{1/2}, & \hbox{Ground state,} \\
                                        8.783\left(\frac{\mu}{\mu^{(1)}_{c}}-1\right)^{1/2}, & \hbox{1st excited state,} \\
                                        14.069\left(\frac{\mu}{\mu^{(2)}_{c}}-1\right)^{1/2}, & \hbox{2st excited state,}
                                      \end{array}
                                    \right.
\end{eqnarray}
where the critical chemical potentials $\mu^{(0)}_{c}$, $\mu^{(1)}_{c}$ and $\mu^{(2)}_{c}$, which correspond to the ground, first and second excited states, are given in Table \ref{SWave} for both operators. Obviously, for both operators $\mathcal{O}_{-}$ and $\mathcal{O}_{+}$ with the excited states, the phase transition between the s-wave holographic insulator and superconductor belongs to the second order and the critical exponent of the system takes the mean-field value $1/2$.

\begin{figure}[H]
\includegraphics[scale=0.65]{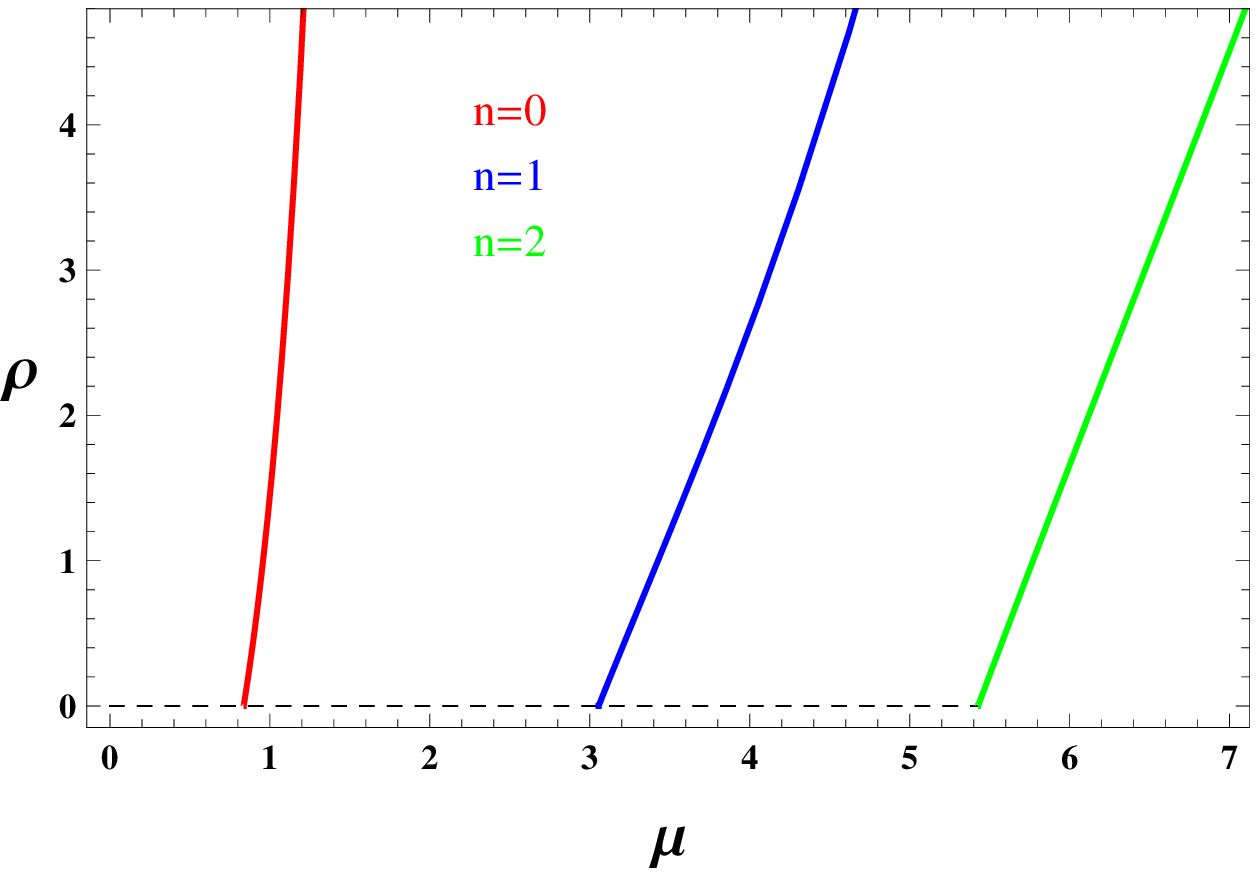}\hspace{0.2cm}%
\includegraphics[scale=0.65]{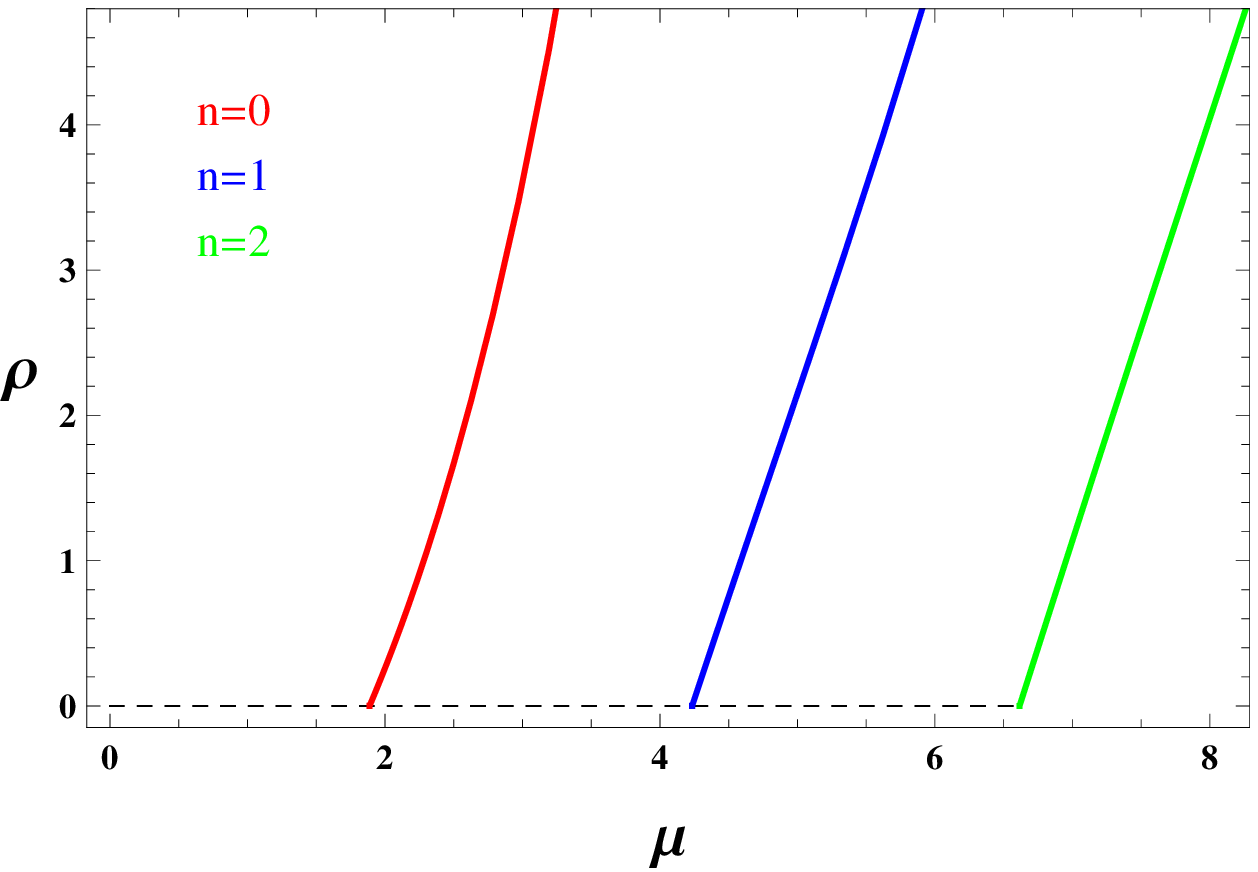}\\ \vspace{0.0cm}
\caption{\label{RealizationSWave} (Color online) The charge density $\rho$ as a function of the chemical potential $\mu$ with fixed mass of the scalar field $m^{2}=-15/4$ when $\langle{\cal
O_{-}}\rangle\neq0$ (left) and $\langle{\cal O_{+}}\rangle\neq0$
(right). In each panel, the three lines from left to right correspond to the ground ($n=0$, red), first ($n=1$, blue) and second ($n=2$, green) excited states, respectively.}
\end{figure}

In order to study the relation between the charge density and chemical potential, in Fig. \ref{RealizationSWave} we exhibit the charge density $\rho$ as a function of the chemical potential with $m^{2}=-15/4$ for the first three lowest-lying modes. For the ground state or each excited state, we observe that the system is described by the AdS soliton solution itself when $\mu$ is small, which can be interpreted as the insulator phase \cite{Nishioka-Ryu-Takayanagi}. But when $\mu\rightarrow\mu^{(n)}_{c}$, there is a phase transition and the AdS soliton reaches the superconductor phase for larger $\mu$. By fitting these curves near the critical point, we find that for the scalar operator $\mathcal{O}_{-}$
\begin{eqnarray}\label{SWaveRhoFu}
\rho\approx \left\{
                                      \begin{array}{ll}
                                        6.863\left(\mu-\mu^{(0)}_{c}\right), & \hbox{Ground state,} \\
                                        2.912\left(\mu-\mu^{(1)}_{c}\right), & \hbox{1st excited state,} \\
                                        2.877\left(\mu-\mu^{(2)}_{c}\right), & \hbox{2st excited state,}
                                      \end{array}
            \right.
\end{eqnarray}
and for the scalar operator $\mathcal{O}_{+}$
\begin{eqnarray}\label{SWaveRhoZheng}
\rho\approx \left\{
                                      \begin{array}{ll}
                                        2.237\left(\mu-\mu^{(0)}_{c}\right), & \hbox{Ground state,} \\
                                        2.846\left(\mu-\mu^{(1)}_{c}\right), & \hbox{1st excited state,} \\
                                        2.976\left(\mu-\mu^{(2)}_{c}\right), & \hbox{2st excited state,}
                                      \end{array}
            \right.
\end{eqnarray}
where again the critical chemical potentials $\mu^{(0)}_{c}$, $\mu^{(1)}_{c}$ and $\mu^{(2)}_{c}$, corresponding to the ground, first and second excited states, are shown in Table \ref{SWave} for both operators $\mathcal{O}_{-}$ and $\mathcal{O}_{+}$. Thus, for both operators $\mathcal{O}_{-}$ and $\mathcal{O}_{+}$, the linear relationship between the charge density and chemical potential $\rho\sim\left(\mu-\mu^{(n)}_{c}\right)$ is valid in general for the s-wave holographic insulator/superconductor phase transition with the excited states in the vicinity of the critical point. The number of nodes $n$ does not affect the observed linearity.

\subsection{Analytical investigation}

In this section, we generalize the Sturm-Liouville method \cite{Siopsis,SiopsisBF} to analytically investigate the properties of the s-wave holographic insulator/superconductor phase transition with the excited states and back up the numerical computations.

\subsubsection{Critical chemical potential}

Introducing a variable $z=r_{s}/r$ for convenience, we can rewrite  Eqs. (\ref{psi}) and (\ref{phi}) into
\begin{eqnarray}
\psi^{\prime\prime}+\left(
\frac{f^\prime}{f}-\frac{1}{z}\right)\psi^\prime
+\left[\frac{1}{z^2f}\left(\frac{q\phi}{r_{s}}\right)^{2}-\frac{m^2}{z^{4}f}\right]\psi=0\,,
\label{psiz}
\end{eqnarray}
\begin{eqnarray}
\phi^{\prime\prime}+\left(\frac{1}{z}+\frac{f^\prime}{f}\right)
\phi^\prime-\frac{2q^{2}\psi^2}{z^4f}\phi=0, \label{phiz}
\end{eqnarray}
where the function $f$ is $f(z)=(1-z^{4})/z^2$ and the prime denotes the derivative with respect to $z$.

It has been shown numerically in the previous section that, for both the ground and excited states of the s-wave holographic dual model in the backgrounds of AdS soliton, there is a phase transition between the insulator and superconductor phases around the critical chemical potential $\mu_{c}$. Since the scalar field $\psi$ vanishes as long as one approaches the critical point $\mu_{c}$ from below, we can get a general solution to Eq. (\ref{phiz}) at the critical point
\begin{eqnarray}\label{phiCriticalPoint}
\phi=\mu+c_{1}\ln\left(\frac{1+z^{2}}{1-z^{2}}\right),
\end{eqnarray}
with an integration constant $c_{1}$. According to the Neumann-like boundary condition for the gauge field $\phi$, we have to set $c_{1}=0$ to keep $A_{t}$ finite at the tip $z=1$. Thus, for the case of $\mu\leq\mu_{c}$, we obtain the physical solution $\phi(z)=\mu$ to Eq. (\ref{phiz}).

In order to match the behavior of $\psi$ at the boundary (\ref{infinity}), we define a trial function $F(z)$ which satisfies
\begin{eqnarray}\label{psiFz}
\psi(z)\simeq\frac{\langle{\cal O}_{i}\rangle}{{r_{s}}^{\Delta_{i}}}z^{\Delta_i}F(z),
\end{eqnarray}
with $i=+/-$. Here we have imposed the boundary condition $F(0)=1$. With the help of Eq. (\ref{psiz}) and the physical solution $\phi(z)$ in Eq. (\ref{phiCriticalPoint}), we find
\begin{eqnarray}\label{Fzmotion}
(TF')'+T\left[U+V{\left(\frac{q\mu}{r_{s}}\right)}^{2}\right]F=0,
\end{eqnarray}
with
\begin{eqnarray}\label{TUV}
T=z^{2\Delta_{i}-1}f,~~~~~~~~~~U=\frac{\Delta_{i}}{z}\left(\frac{\Delta_{i}-2}{z}+\frac{f'}{f}\right)-\frac{m^{2}}{z^{4}f},
~~~~~~~~~~V=\frac{1}{z^{2}f}.
\end{eqnarray}
According to the Sturm-Liouville eigenvalue approach \cite{Gelfand-Fomin}, the eigenvalue $q\mu/r_{s}$ can be achieved from the extremal values of the following function by virtue of the Rayleigh Quotient
\begin{eqnarray}\label{EigenvalueSWave}
\left(\frac{q\mu}{r_{s}}\right)^{2}=\frac{\int^{1}_{0}T\left(F'^{2}-UF^{2}\right)dz}{\int^{1}_{0}TVF^{2}dz}.
\end{eqnarray}

Before proceeding, we would like to make a comment. In order to derive the expression (\ref{EigenvalueSWave}), we have employed the boundary condition $[T(z)F(z)F'(z)]|_{0}^{1}=0$. Note that $T(1)\equiv0$ from Eq. (\ref{TUV}), so the condition $T(1)F(1)F'(1)=0$ is satisfied automatically. However, for the case of $m^{2}=-15/4$ considered here, the condition $T(0)F(0)F'(0)=0$ is not satisfied automatically for the operator $\mathcal{O}_{-}$ since the leading order contribution from $T(z)$ as $z\rightarrow0$ is $2\Delta_{-}-3=0$ but is satisfied automatically for the operator $\mathcal{O}_{+}$ since $2\Delta_{+}-3=2>0$. Thus, we have to impose the Neumann boundary condition $F'(0)=0$ for the operator $\mathcal{O}_{-}$, just as analyzed in Refs. \cite{QiaoEHS,LvPLB2020,HFLi,WangSPJ}.

As an example, we calculate the case for the operator $\mathcal{O}_{-}$ by using the fourth order trial function
\begin{eqnarray}\label{SWTrialFunction}
F(z)=1-a_{2}z^{2}-a_{3}z^{3}-a_{4}z^{4},
\end{eqnarray}
which satisfies the Neumann boundary condition $F'(0)=0$ with three constants $a_{2}$, $a_{3}$ and $a_{4}$. From Eq. (18), we have
\begin{eqnarray}
\begin{split}
\left(\frac{q\mu}{r_{s}}\right)^{2}=\left(\frac{3}{4}-\frac{9 a_2}{10}-\frac{3 a_3}{4}-\frac{9 a_4}{14}+\frac{33 a_2 a_3}{16}+\frac{173 a_2 a_4}{90}+\frac{41 a_3 a_4}{20}+\frac{13 a_2^2}{12}+\frac{21 a_3^2}{20}
+\frac{29 a_4^2}{28}\right)/\\
\left(1-\frac{2 a_2}{3}-\frac{a_3}{2}-\frac{2 a_4}{5}+\frac{a_2 a_3}{3}+\frac{2 a_2 a_4}{7}+\frac{a_3 a_4}{4}+\frac{a_2^2}{5}+\frac{a_3^2}{7}+\frac{a_4^2}{9}\right),
\end{split}
\end{eqnarray}
which gives us the dimensionless critical chemical potential $q\mu_{c}/r_{s}$ and corresponding value of $a_{k}$ from the ground state to the third excited state by computing the extremal values of the above expression, i.e., $q\mu_{c}^{(0)}/r_{s}=0.836$ at $a_{2}=0.384$, $a_{3}=-0.157$ and $a_{4}=-0.002$, $q\mu_{c}^{(1)}/r_{s}=3.053$ at $a_{2}=5.298$, $a_{3}=-2.998$ and $a_{4}=0.231$, $q\mu_{c}^{(2)}/r_{s}=5.432$ at $a_{2}=20.306$, $a_{3}=-30.120$ and $a_{4}=8.793$, and $q\mu_{c}^{(3)}/r_{s}=7.847$ at $a_{2}=47.655$, $a_{3}=-120.749$ and $a_{4}=76.511$. Comparing with the analytical results from the second order trial function in \cite{HFLi}, we can obtain the first four lowest-lying modes by using the fourth order trial function $F(z)$.

In order to investigate the higher excited states of the s-wave holographic insulator/superconductor phase transition by using the analytical Sturm-Liouville method, we include the eighth order of $z$ in the trial function $F(z)$, i.e., $F(z)=1-\sum_{k=2}^{k=8}a_{k}z^{k}$ for the operator $\mathcal{O}_{-}$, and $F(z)=1-\sum_{k=1}^{k=8}a_{k}z^{k}$ for the operator $\mathcal{O}_{+}$ in the following calculation, where $a_{k}$ is a constant. Using the expression (\ref{EigenvalueSWave}) to compute the extremal values, we can get the critical chemical potentials from the ground state to the sixth excited state, which have been presented in Tables \ref{SWaveO1Table} and \ref{SWaveO2Table}. Compared with the numerical results shown in Table \ref{SWave}, the agreement of the analytical results and numerical calculation is impressive, which implies that the Sturm-Liouville method is powerful to study the s-wave holographic insulator/superconductor phase transition even if we consider the excited states.
\begin{table}[ht]
\begin{center}
\caption{\label{SWaveO1Table}
The dimensionless critical chemical potential $q\mu_{c}/r_{s}$ obtained by the Sturm-Liouville method for the operator $\mathcal{O}_{-}$ and corresponding value of $a_{k}$ for the trial function $F(z)=1-\sum_{k=2}^{k=8}a_{k}z^{k}$ from the ground state to the sixth excited state in the s-wave holographic insulator and superconductor model. Here we fix the mass of the scalar field by $m^{2}=-15/4$.}
\begin{tabular}{c |c| c c c c c c c c }
\hline
~$n$~&~$q\mu_{c}/r_{s}$~&~$a_{2}$~&~$a_{3}$~&~$a_{4}$~&~$a_{5}$~&~$a_{6}$~&~$a_{7}$~&~$a_{8}$~  \\
\hline
~$0$~ &~0.836~~&~0.349~~&~0.001~~&~-0.212~~&~-0.006~~&~0.212~~&~-0.158~~&~0.038~ \\
\hline
~$1$~ &~3.053~~&~4.657~~&~0.035~~&~-3.951~~&~0.054~~&~4.230~~&~-3.306~~&~0.826~ \\
\hline
~$2$~ &~5.423~~&~14.643~~&~0.909~~&~-40.778~~&~8.181~~&~47.340~~&~-43.031~~&~11.673~ \\
\hline
~$3$~ &~7.810~~&~29.518~~&~15.289~~&~-241.971~~&~217.724~~&~155.313~~&~-258.306~~&~85.914~ \\
\hline
~$4$~ &~10.202~~&~42.928~~&~153.621~~&~-1405.255~~&~2784.640~~&~-2025.254~~&~347.735~~&~99.762~\\
\hline
~$5$~ &~12.603~~&~37.159~~&~785.173~~&~-6433.838~~&~17725.045~~&~-22568.526~~&~13453.361~~&~-2294.129~ \\
\hline
~$6$~ &~15.040~~&~68.959~~&~1102.374~~&~-12006.310~~&~41434.653~~&~-66302.827~~&~50620.592~~&~-14920.513~ \\
\hline
\end{tabular}
\end{center}
\end{table}

\begin{table}[ht]
\begin{center}
\caption{\label{SWaveO2Table}
The dimensionless critical chemical potential $q\mu_{c}/r_{s}$ obtained by the Sturm-Liouville method for the operator $\mathcal{O}_{+}$ and corresponding value of $a_{k}$ for the trial function $F(z)=1-\sum_{k=1}^{k=8}a_{k}z^{k}$ from the ground state to the sixth excited state in the s-wave holographic insulator and superconductor model. Here we fix the mass of the scalar field by $m^{2}=-15/4$.}
\begin{tabular}{c |c| c c c c c c c c c}
\hline
~$n$~&~$q\mu_{c}/r_{s}$~&~$a_{1}$~&~$a_{2}$~&~$a_{3}$~&~$a_{4}$~&~$a_{5}$~&~$a_{6}$~&~$a_{7}$~&~$a_{8}$~ \\
\hline
~$0$~ &~1.888~~&~0.000(3)~~&~0.590~~&~0.030~~&~-0.520~~&~0.167~~&~0.267~~&~-0.240~~&~0.060~ \\
\hline
~$1$~ &~4.234~~&~0.003~~&~2.941~~&~0.317~~&~-4.102~~&~1.983~~&~1.584~~&~-1.774~~&~0.479~ \\
\hline
~$2$~ &~6.616&~~0.033~~&~6.791~~&~3.452~~&~-28.924~~&~24.753~~&~-0.042~~&~-8.248~~&~2.840~ \\
\hline
~$3$~ &~9.005~~&~0.263~~&~9.385~~&~29.248~~&~-167.855~~&~240.706~~&~-137.001~~&~23.744~~&~2.803~ \\
\hline
~$4$~ &~11.398~~&~1.289~~&~0.108~~&~160.370~~&~-800.342~~&~1541.875~~&~-1429.746~~&~637.510~~&~-110.319~ \\
\hline
~$5$~ &~13.792~~&~2.422~~&~-15.760~~&~410.605~~&~-2258.564~~&~5315.877~~&~-6254.036~~&~3618.816~~&~-818.134~ \\
\hline
~$6$~ &~16.263~~&~-9.719~~&~175.378~~&~-559.660~~&~-315.936~~&~4517.559~~&~-8679.681~~&~6911.895~~&~-2039.153~ \\
\hline
\end{tabular}
\end{center}
\end{table}

From Tables \ref{SWaveO1Table} and \ref{SWaveO2Table}, we confirm analytically that for both operators the critical chemical potential $\mu_{c}$ increases as the number of nodes $n$ increases, which supports the numerical observation obtained in Table \ref{SWave} that an excited state has a higher critical chemical potential than the corresponding ground state. Fitting the relation between $q\mu_{c}/r_{s}$ and $n$ by using the analytical results, we get
\begin{eqnarray}\label{SWaveMucAna}
\frac{q\mu_{c}}{r_{s}}\approx
\left\{
\begin{array}{rl}
2.375n+0.728, &  \quad {\rm for} \ \mathcal{O}_{-}\,,\\ \\
2.394n+1.847, &  \quad {\rm for} \  \mathcal{O}_{+}\,,
\end{array}\right.
\end{eqnarray}
which is in good agreement with the numerical fitting results given in Eq. (\ref{SWaveMuc}). This can be used to back up the numerical computation that, for both operators $\mathcal{O}_{-}$ and $\mathcal{O}_{+}$, the difference of the dimensionless critical chemical potential $q\mu_{c}/r_{s}$ between the consecutive states is about 2.4.

\begin{figure}[H]
\includegraphics[scale=0.65]{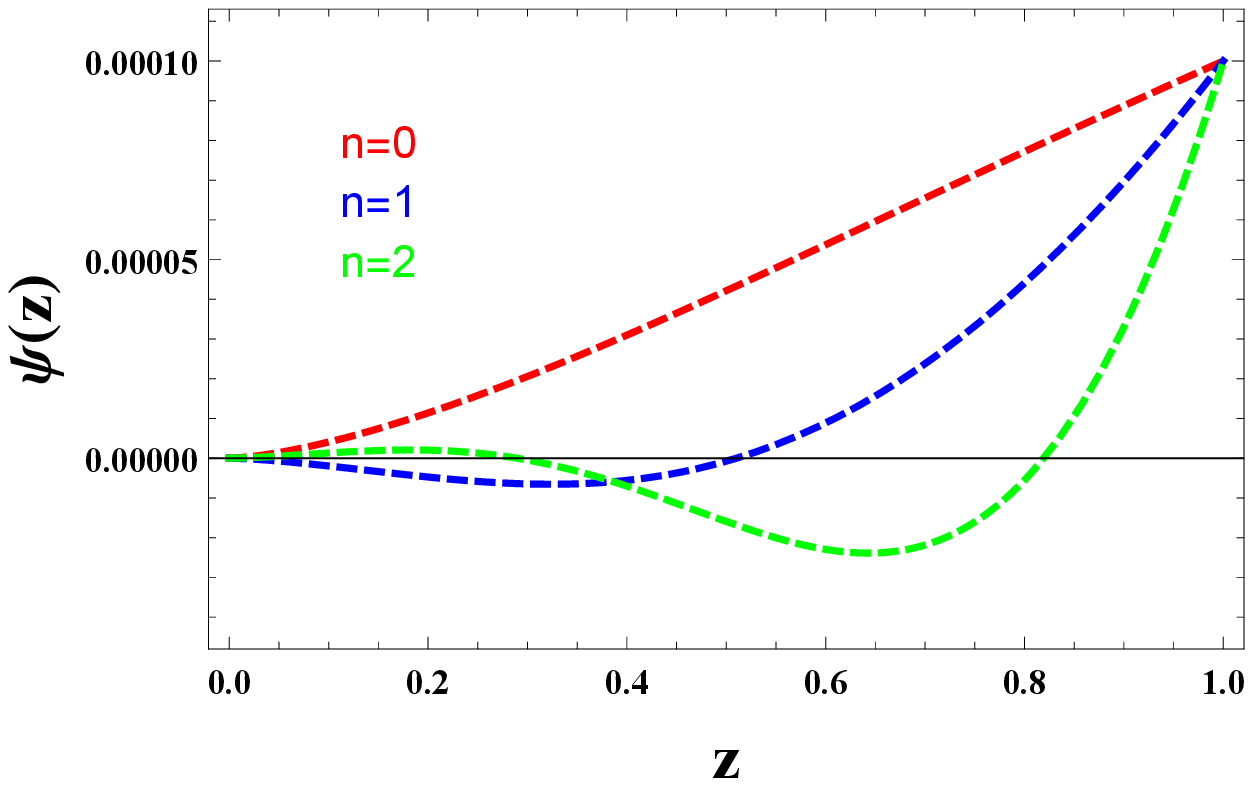}\hspace{0.2cm}%
\includegraphics[scale=0.65]{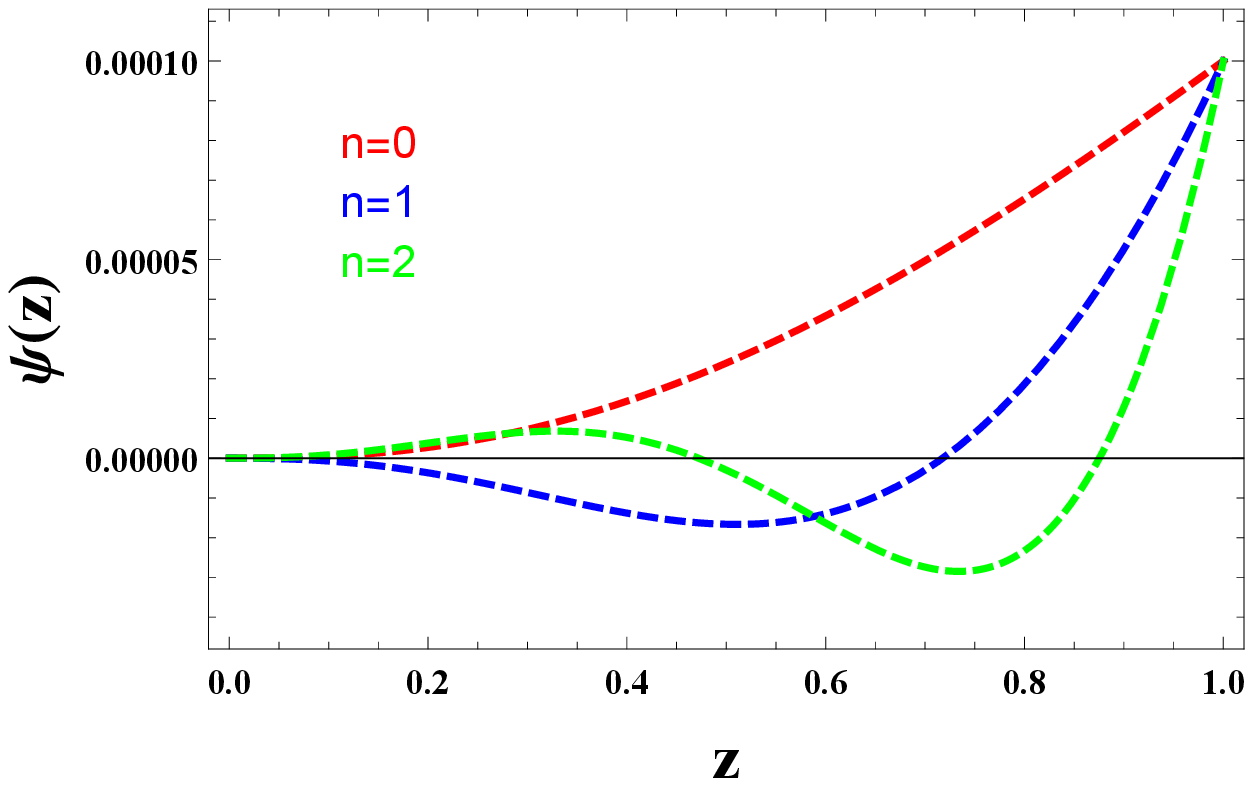}\\ \vspace{0.0cm}
\caption{\label{PsizAna} (Color online) The scalar field $\psi(z)$ as a function of the radial coordinate $z$ outside the horizon with the scalar operators $\mathcal{O}_{-}$ (left) and $\mathcal{O}_{+}$ (right) for the fixed mass of the scalar field $m^{2}=-15/4$ by using the analytical Sturm-Liouville method with $\psi(1)=0.0001$ in Eq. (\ref{psiFz}). In each panel, the three lines from top to bottom correspond to the ground ($n=0$, red), first ($n=1$, blue) and second ($n=2$, green) excited states, respectively.}
\end{figure}

On the other hand, for the scalar field mass $m^{2}=-15/4$, in Fig. \ref{PsizAna} we use the expression (\ref{psiFz}) to give the distribution of the scalar field $\psi(z)$ as a function of $z$ for the scalar operators $\mathcal{O}_{-}$ and $\mathcal{O}_{+}$ by setting the initial condition $\psi(1)=0.0001$, which agrees well with the numerical calculation shown in Fig. \ref{Psiz}. Interestingly enough, besides the critical chemical potential, the Sturm-Liouville method with the higher order of $z$ in the trial function $F(z)$ can also study the behaviors of the scalar field near the critical point of the phase transition.

\subsubsection{Critical phenomena}

Since the condensate of the scalar operator $\langle{\cal O}_{i}\rangle$ is so small near the critical point, we can expand $\phi(z)$ in terms of small $\langle{\cal O}_{i}\rangle$ by
\begin{eqnarray}
\phi(z)\sim\mu_{c}+2\mu_{c}\left(\frac{q\langle{\cal O}_{i}\rangle}{{r_{s}}^{\Delta_{i}}}\right)^{2}\chi(z)+\cdots,
\end{eqnarray}
which leads to the equation of motion for $\chi(z)$
\begin{eqnarray}\label{SWChi}
(W\chi')'-z^{2\Delta_{i}-3}F^{2}=0,
\end{eqnarray}
where we have defined a new function $W(z)=zf(z)$ and introduced the boundary condition $\chi(1)=0$ at the
tip.

Considering the asymptotic behavior of $\phi$ in Eq. (\ref{infinity}), we expand $\phi$ when $z\rightarrow 0$ as
\begin{eqnarray}\label{ExpandingSWphi}
\phi\simeq\mu-\frac{\rho }{r_{s}^{2}}z^{2}\simeq\mu_{c}+2\mu_{c}\left(\frac{q\langle{\cal
O}_{i}\rangle}{r_{s}^{\Delta_{i}}}\right)^{2}\left[\chi(0)+\chi'(0)z+\frac{1}{2}\chi''(0)z^{2}+\cdots\right].
\end{eqnarray}
Comparing the coefficients of the $z^{0}$ term in both sides of the above formula, we can easily get
\begin{eqnarray}\label{SWaveOi}
\frac{q\langle{\cal
O}_{i}\rangle}{r_{s}^{\Delta_{i}}}=\frac{1}{\sqrt{2\chi(0)}}\left(\frac{\mu}{\mu_{c}}-1\right)^{1/2},
\end{eqnarray}
where $\chi(0)=c_{2}-\int_{0}^{1}W^{-1}[c_{3}+\int_{1}^{z}x^{2\Delta_{i}-3}F(x)^{2}dx]dz$ with the integration constants $c_{2}$ and $c_{3}$ which can be determined by the boundary condition of $\chi(z)$ in Eq. (\ref{SWChi}). For example, in the case of $m^{2}=-15/4$ considered here, for the operator $\mathcal{O}_{-}$ we find that
\begin{eqnarray}
\frac{q\langle{\cal O}_{-}\rangle}{r_{s}^{3/2}}\approx \left\{
                                      \begin{array}{ll}
                                        1.799\left(\frac{\mu}{\mu^{(0)}_{c}}-1\right)^{1/2}, & \hbox{Ground state,} \\
                                        1.396\left(\frac{\mu}{\mu^{(1)}_{c}}-1\right)^{1/2}, & \hbox{1st excited state,} \\
                                        1.394\left(\frac{\mu}{\mu^{(2)}_{c}}-1\right)^{1/2}, & \hbox{2st excited state,}
                                      \end{array}
                                    \right.
\end{eqnarray}
which agrees well with the numerical result given in Eq. (\ref{SWaveOFu}). Here the critical chemical potentials $\mu^{(0)}_{c}$, $\mu^{(1)}_{c}$ and $\mu^{(2)}_{c}$, which correspond to the ground, first and second excited states, are given in Table \ref{SWaveO1Table}. Obviously, the expression (\ref{SWaveOi}) is valid for all cases considered here. Therefore, it is shown clearly that, for both operators $\mathcal{O}_{-}$ and $\mathcal{O}_{+}$, the phase transition between the s-wave insulator and superconductor is second order and the condensate approaches zero as $\langle{\cal O}_{i}\rangle\sim\left(\mu-\mu^{(n)}_{c}\right)^{\beta}$ with the mean-field critical exponent $\beta=1/2$ for all the states.

From the coefficients of the $z^{1}$ term in Eq. (\ref{ExpandingSWphi}), we note that $\chi'(0)\rightarrow 0$, which is consistent with the following relation by making integration of both sides of Eq. (\ref{SWChi})
\begin{eqnarray}
\left[\frac{\chi'(z)}{z}\right]\bigg|_{z\rightarrow0}=\chi''(0)=-\int_{0}^{1}z^{2\Delta_{i}-3}F^{2}dz.
\end{eqnarray}

Moving to the coefficients of the $z^{2}$ term in Eq. (\ref{ExpandingSWphi}), we obtain
\begin{eqnarray}\label{SWaveRhoAna}
\frac{\rho}{r_{s}^{2}}=-\mu_{c}\chi''(0)\left(\frac{q\langle{\cal
O}_{i}\rangle}{r_{s}^{\Delta_{i}}}\right)^{2}=\Gamma(m,n)(\mu-\mu_{c}),
\end{eqnarray}
with $\Gamma(m,n)=[2\chi(0)]^{-1}\int_{0}^{1}z^{2\Delta_{i}-3}F^{2}dz$, which is a function of the scalar field mass $m^{2}$ and the number of nodes $n$. As an example, we observe that for the operator $\mathcal{O}_{-}$ with the fixed mass $m^{2}=-15/4$
\begin{eqnarray}
\frac{\rho}{r_{s}^{2}}\approx \left\{
                                      \begin{array}{ll}
                                        2.707\left(\mu-\mu^{(0)}_{c}\right), & \hbox{Ground state,} \\
                                        1.301\left(\mu-\mu^{(1)}_{c}\right), & \hbox{1st excited state,} \\
                                        1.281\left(\mu-\mu^{(2)}_{c}\right), & \hbox{2st excited state,}
                                      \end{array}
                                    \right.
\end{eqnarray}
which again can be compared with the numerical result presented in Eq. (\ref{SWaveRhoFu}), where the critical chemical potentials $\mu^{(0)}_{c}$, $\mu^{(1)}_{c}$ and $\mu^{(2)}_{c}$ are given in Table \ref{SWaveO1Table}, corresponding to the ground, first and second excited states, respectively. Thus, we point out that, in the vicinity of the transition point, one may find a linear relationship between the charge density and chemical potential, namely, $\rho\sim\left(\mu-\mu^{(n)}_{c}\right)$ in the present model for all the states, which is in agreement with the numerical calculation shown previously in Eqs. (\ref{SWaveRhoFu}) and (\ref{SWaveRhoZheng}) for both operators $\mathcal{O}_{-}$ and $\mathcal{O}_{+}$.

\section{Excited states of the p-wave holographic insulator/superconductor phase transition}

In the previous section, we have investigated the excited states of the s-wave holographic insulator/superconductor phase transition by employing the numerical shooting method as well as the analytical Sturm-Liouville approach. Now, we extend our study to the excited states of the p-wave holographic insulator/superconductor phase transition in the probe limit by considering the Maxwell complex vector field model \cite{CaiPWave-1,CaiPWave-2}
\begin{eqnarray}\label{PWaveSystem}
S=\int d^{5}x\sqrt{-g}\left(-\frac{1}{4}F_{\mu\nu}F^{\mu\nu}-\frac{1}{2}\rho_{\mu\nu}^{\dag}\rho^{\mu\nu}-m^{2}\rho_{\mu}^{\dag}\rho^{\mu}+iq\gamma\rho_{\mu}\rho_{\nu}^{\dag}F^{\mu\nu}\right),
\end{eqnarray}
where the tensor $\rho_{\mu\nu}$ is defined by $\rho_{\mu\nu}=(\nabla_{\mu}-iqA_{\mu})\rho_{\nu}-(\nabla_{\nu}-iqA_{\nu})\rho_{\mu}$, the parameter $\gamma$ describes the interaction between the vector field $\rho_\mu$ and the gauge field $A_\mu$, $q$ and $m$ represent the charge and mass of the vector field $\rho_{\mu}$. Making use of the ansatz for the matter fields $\rho_{\mu}dx^{\mu}=\rho_{x}(r)dx$ and $A_\mu dx^{\mu}=A_t(r)dt$, we obtain the equations of motion
\begin{eqnarray}\label{PWaveRhox}
\rho_{x}''+\left(\frac{1}{r}+\frac{f'}{f}\right)\rho_{x}'-\frac{1}{f}\left(m^{2}-\frac{q^{2}A_{t}^{2}}{r^{2}}\right)\rho_{x}=0,
\end{eqnarray}
\begin{eqnarray}\label{PWaveAt}
A_{t}''+\left(\frac{1}{r}+\frac{f'}{f}\right)A_{t}'-\frac{2q^{2}\rho_{x}^{2}}{r^{2}f}A_{t}=0,
\end{eqnarray}
where the prime denotes the derivative with respect to $r$. It should be noted that, if we set $m^{2}=0$, $A_t=\phi$ and rescale the vector field by $\rho_{x}=\psi/\sqrt{2}$, we can easily recover the equations of motion (4.4) and (4.5) in \cite{HFLi} for the p-wave holographic insulator/superconductor phase transition where an $SU(2)$ Yang-Mills action is considered.

In order to solve Eqs. (\ref{PWaveRhox}) and (\ref{PWaveAt}), the vector field $\rho_\mu$ and gauge field $A_\mu$ are required to be regular at the tip $r=r_{s}$. And as $r\rightarrow\infty$, the asymptotical behaviors are
\begin{eqnarray}
\rho_{x}=\frac{\rho_{x_{-}}}{r^{\Delta_{-}}}+\frac{\rho_{x_{+}}}{r^{\Delta_{+}}}\,,\hspace{0.5cm}
A_{t}=\mu-\frac{\rho}{r^{2}}\,, \label{PWaveinfinity}
\end{eqnarray}
where $\Delta_{\pm}=1\pm\sqrt{1+m^{2}}$ are the characteristic exponents, and $\rho_{x-}$ and $\rho_{x+}$ are interpreted as the source and the vacuum expectation value of the vector operator $ \langle{\cal O}_{x}\rangle$ in the dual field theory from the AdS/CFT correspondence \cite{NieHZ,HuangSCPMA}, respectively. In this work, we impose boundary condition $\rho_{x_{-}}=0$ since we are interested in the case where the condensate appears spontaneously. And we use $\Delta$ to denote $\Delta_{+}$ for simplicity and set $m^{2}=5/4$ for concreteness. In order to compare our results with those in the p-wave holographic insulator/superconductor model in the Yang-Mills theory \cite{CaiLZSolition,HFLi}, we also present the results for the case $m^{2}=0$ simultaneously in the following.

\subsection{Numerical analysis}

From the equations of motion (\ref{PWaveRhox}) and (\ref{PWaveAt}), one may find that the following scaling symmetries
\begin{eqnarray}
&&r\rightarrow \lambda r,~~~(t,\varphi,x,y)\rightarrow\frac{1}{\lambda}(t,\varphi,x,y),~~~q\rightarrow q,~~~(\rho_{x},A_{t})\rightarrow\lambda(\rho_{x},A_{t}),\nonumber \\
&&\mu\rightarrow\lambda\mu,~~~\rho\rightarrow\lambda^{3}\rho,~~~~~~\rho_{x_{+}}\rightarrow\lambda^{1+\Delta}\rho_{x_{+}},
\end{eqnarray}
with a real positive number $\lambda$, are held. Therefore, we can use them to choose $r_{s}=1$ and $q=1$ throughout the numerical calculations, similar to the study shown in the last section for the s-wave holographic insulator/superconductor model.

\begin{figure}[H]
\includegraphics[scale=0.65]{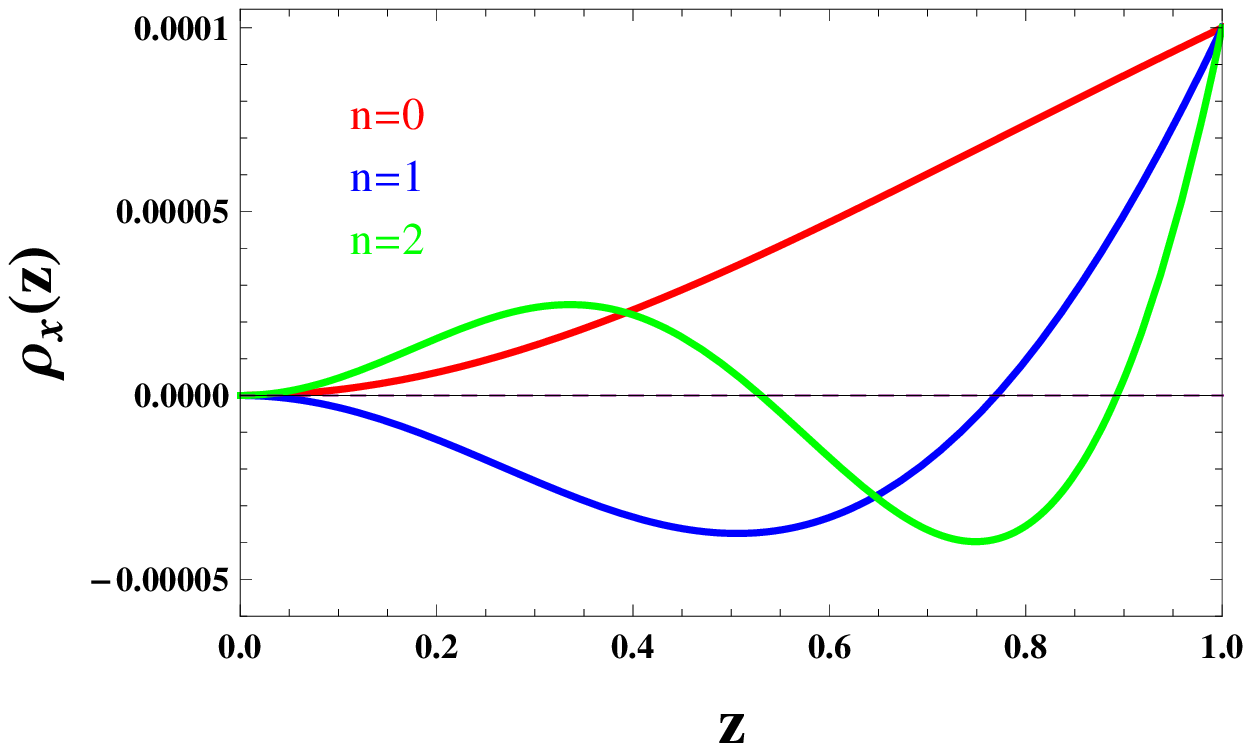}\hspace{0.2cm}%
\includegraphics[scale=0.65]{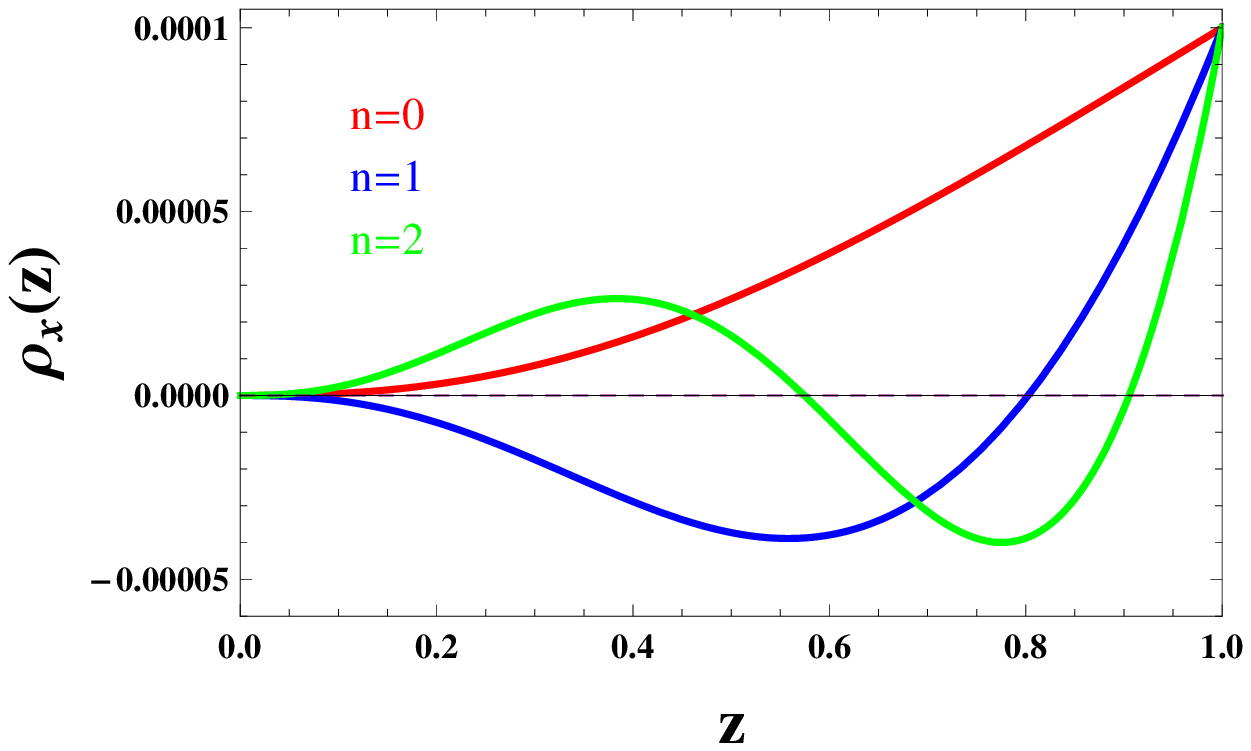}\\ \vspace{0.0cm}
\caption{\label{PWRhoxz} (Color online) The vector field $\rho_{x}(z)$ as a function of the radial coordinate $z$ outside the horizon with the vector operator $\mathcal{O}_{x}$ for the fixed masses of the vector field $m^{2}=0$ (left) and $m^{2}=5/4$ (right) by using the numerical shooting method. In each panel, the three lines from top to bottom correspond to the ground ($n=0$, red), first ($n=1$, blue) and second ($n=2$, green) excited states, respectively.}
\end{figure}

In Fig. \ref{PWRhoxz}, we show the distribution of the vector field $\rho_{x}(z)$ as a function of $z$ for the vector operator $\mathcal{O}_{x}$ with the fixed masses of the vector field $m^{2}=0$ and $m^{2}=5/4$. In each panel, similar to the scalar field $\psi(z)$ in the s-wave holographic insulator/superconductor model, the red line, blue line and green line of the vector field $\rho_{x}(z)$ correspond to the ground state with $n=0$, first excited state with $n=1$ and second excited with $n=2$, respectively. This means that, for both the s-wave and p-wave holographic models, there exist exactly $n$ nodes in the $n$-th excited state.

\begin{table}[ht]
\caption{\label{PWave} The critical chemical potential $\mu_{c}$ obtained by the shooting method for the vector operator $\mathcal{O}_{x}$ with the fixed masses of the vector field $m^{2}=0$ and $m^{2}=5/4$ from the ground state to the sixth excited state.}
\begin{tabular}{c c c c c c c c}
         \hline
$n$ & 0 & 1 & 2 & 3 & 4 & 5 & 6
        \\
        \hline
~~~~$m^{2}=0$~~~~&~~~~~$2.265$~~~~~&~~~~~$4.741$~~~~~&~~~~~$7.156$~~~~~&~~~~~$9.561$~~~~
&~~~~~$11.962$~~~~&~~~~~$14.362$~~~~&~~~~~$16.760$~~~~
          \\
~~~~$m^{2}=5/4$~~~~&~~~~~$2.785$~~~~~&~~~~~$5.291$~~~~~&~~~~~$7.720$~~~~~&~~~~~$10.133$~~~~
&~~~~~$12.539$~~~~&~~~~~$14.943$~~~~&~~~~~$17.344$~~~~
          \\
        \hline
\end{tabular}
\end{table}

In Table \ref{PWave}, we present the critical chemical potential $\mu_{c}$ obtained by the shooting method with the fixed masses of the vector field $m^{2}=0$ and $m^{2}=5/4$ from the ground state to the sixth excited state for the vector operator $\mathcal{O}_{x}$, which shows that, regardless of the vector field mass, the critical chemical potential $\mu_{c}$ increases as the number of nodes $n$ increases. This behavior is reminiscent of that observed for the s-wave holographic insulator/superconductor case, so we conclude that an excited state has a higher critical chemical potential than the corresponding ground state. Fitting these numerical results for the vector operator $\mathcal{O}_{x}$, we have
\begin{eqnarray}\label{PWaveMuc}
\mu_{c}\approx
\left\{
\begin{array}{rl}
2.412n+2.308, &  \quad {\rm for} \  m^{2}=0\,,\\ \\
2.421n+2.844, &  \quad {\rm for} \  m^{2}=5/4\,.
\end{array}\right.
\end{eqnarray}
Compared with the numerical results in Eq. (\ref{SWaveMuc}) for the s-wave holographic model, it is clear that, although the underlying mechanism remains mysterious, the difference of the dimensionless critical chemical potential $\mu_{c}$ between the consecutive states is about 2.4 for both the s-wave and p-wave holographic insulator/superconductor phase transitions.

\begin{figure}[H]
\includegraphics[scale=0.65]{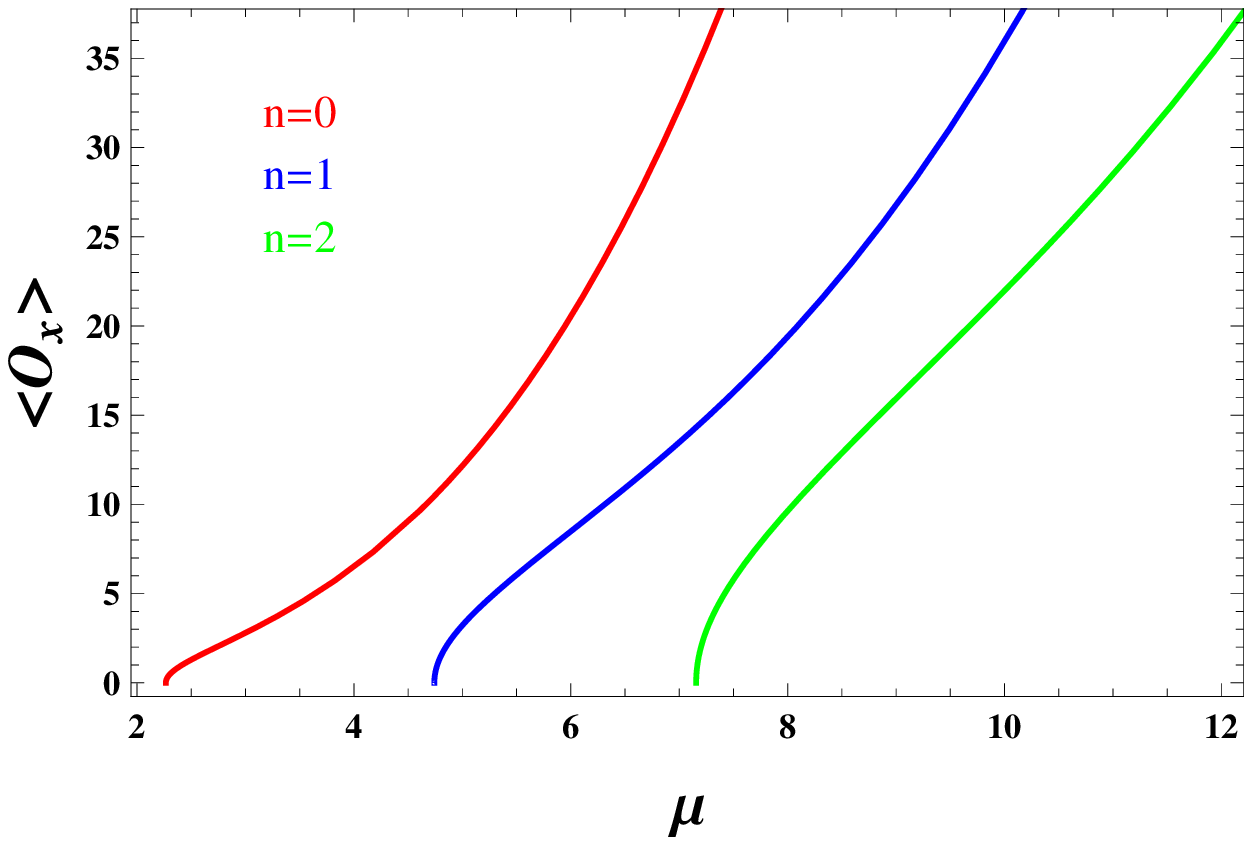}\hspace{0.2cm}%
\includegraphics[scale=0.65]{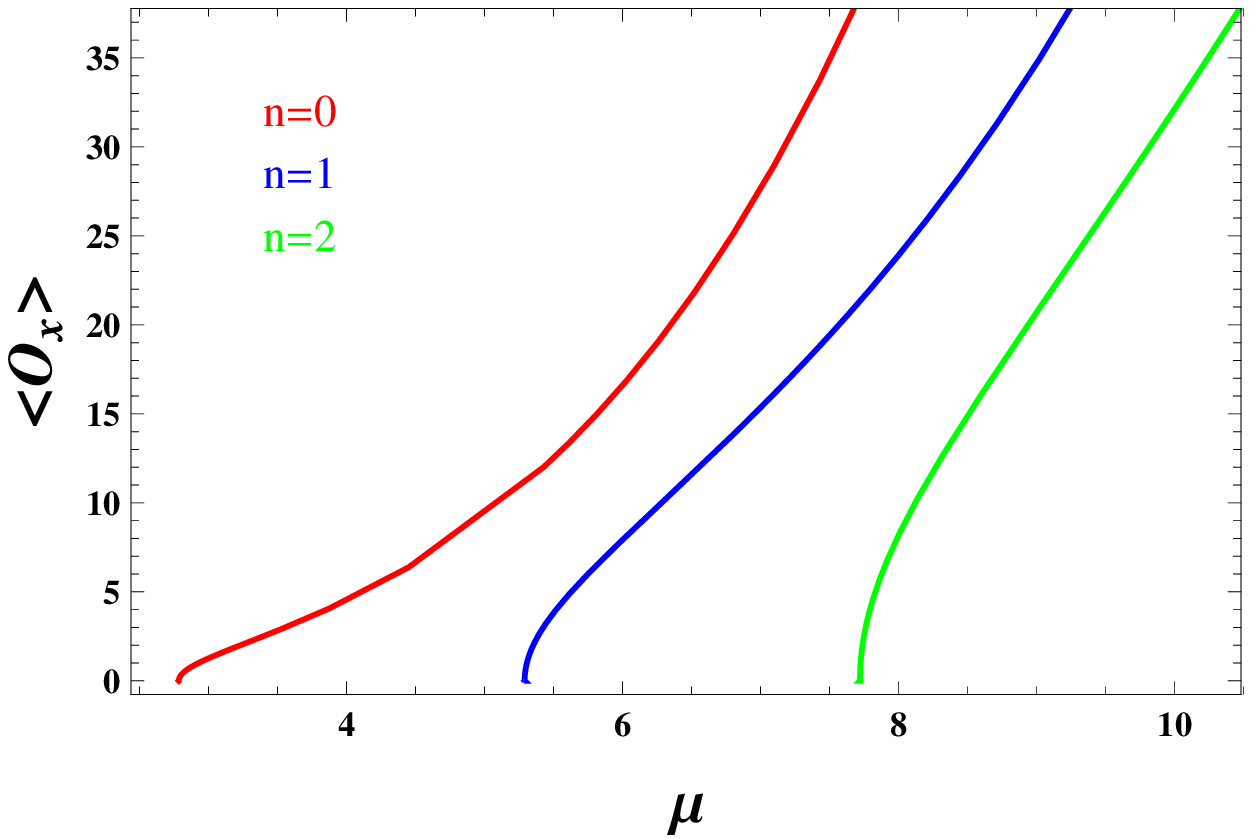}\\ \vspace{0.0cm}
\caption{\label{CondPWave} (Color online) The condensates of
the vector operator $\mathcal{O}_{x}$ with respect to the chemical potential $\mu$ for the masses of the vector field $m^{2}=0$ (left) and $m^{2}=5/4$ (right). In each panel, the three lines from left to right correspond to the ground ($n=0$, red), first ($n=1$, blue) and second ($n=2$, green) excited states, respectively.}
\end{figure}

In Fig. \ref{CondPWave}, we exhibit the condensates of the vector operator $\mathcal{O}_{x}$ as a function of the chemical potential with the masses of the vector field $m^{2}=0$ and $m^{2}=5/4$ for the first three lowest-lying modes $n=0$, $1$ and $2$. Similar to the behavior of the ground state in the probe limit \cite{AkhavanPWaveSolition}, for an excited
state the phase transition occurs as the chemical potential is over a critical value $\mu_{c}$, which can be used to describe the p-wave phase transition between the insulator and superconductor with the excited state. By fitting these curves near $\mu_{c}$, we find that for the vector field mass $m^{2}=0$
\begin{eqnarray}\label{PWaveOxNM0}
\langle{\cal O}_{x}\rangle\approx \left\{
                                      \begin{array}{ll}
                                        3.664\left(\frac{\mu}{\mu^{(0)}_{c}}-1\right)^{1/2}, & \hbox{Ground state,} \\
                                        13.054\left(\frac{\mu}{\mu^{(1)}_{c}}-1\right)^{1/2}, & \hbox{1st excited state,} \\
                                        25.318\left(\frac{\mu}{\mu^{(2)}_{c}}-1\right)^{1/2}, & \hbox{2st excited state,}
                                      \end{array}
                                    \right.
\end{eqnarray}
and for the vector field mass $m^{2}=5/4$
\begin{eqnarray}\label{PWaveOxNM125}
\langle{\cal O}_{x}\rangle\approx \left\{
                                      \begin{array}{ll}
                                        4.191\left(\frac{\mu}{\mu^{(0)}_{c}}-1\right)^{1/2}, & \hbox{Ground state,} \\
                                        18.251\left(\frac{\mu}{\mu^{(1)}_{c}}-1\right)^{1/2}, & \hbox{1st excited state,} \\
                                        41.339\left(\frac{\mu}{\mu^{(2)}_{c}}-1\right)^{1/2}, & \hbox{2st excited state,}
                                      \end{array}
                                    \right.
\end{eqnarray}
where the critical chemical potentials $\mu^{(0)}_{c}$, $\mu^{(1)}_{c}$ and $\mu^{(2)}_{c}$ are given in Table \ref{PWave} for the masses of the vector field $m^{2}=0$ and $m^{2}=5/4$, corresponding to the ground, first and second excited states, respectively. Thus, similar to the s-wave holographic model, the p-wave holographic insulator/superconductor phase transition in the excited states is always the second order with the mean-field critical exponent $1/2$, and the number of nodes $n$ does not affect the order of phase transition.

\begin{figure}[H]
\includegraphics[scale=0.65]{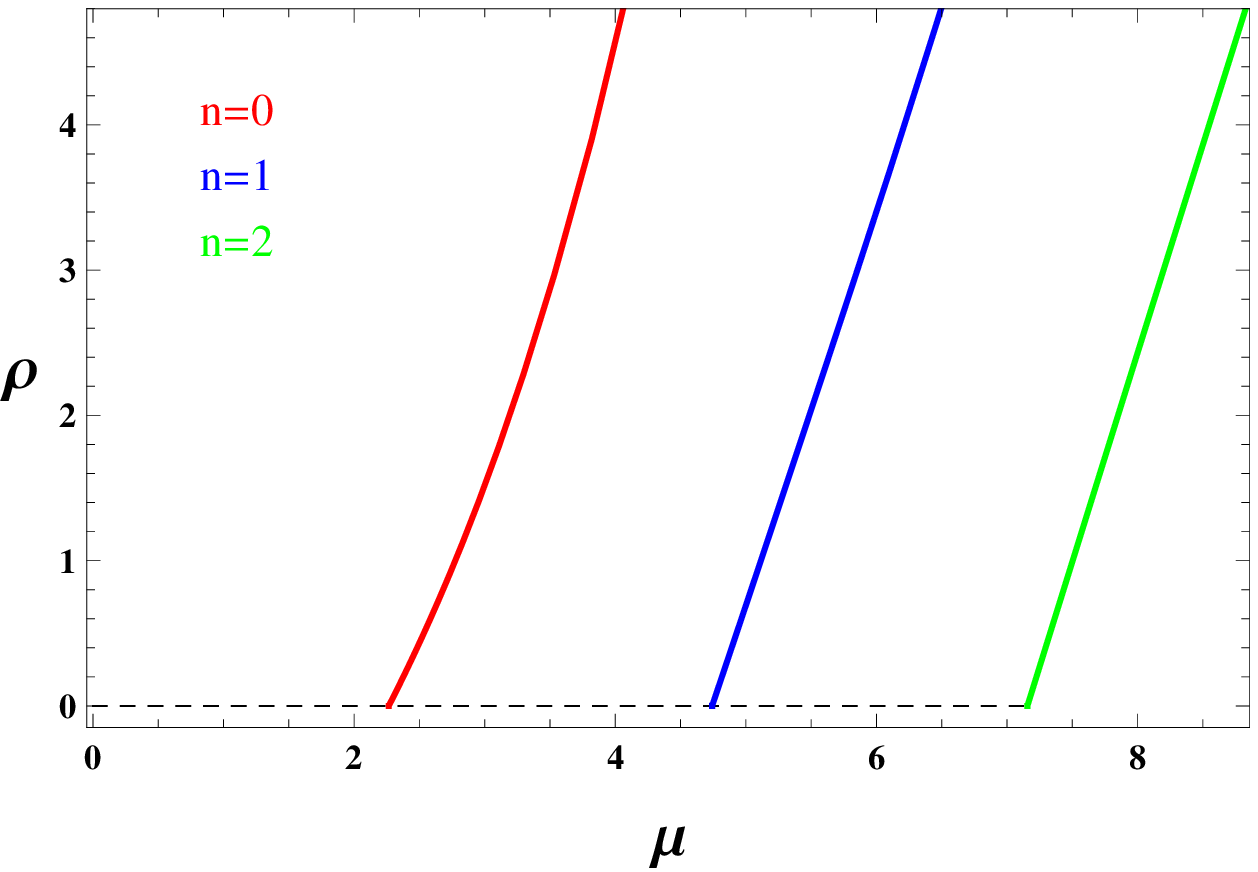}\hspace{0.2cm}%
\includegraphics[scale=0.65]{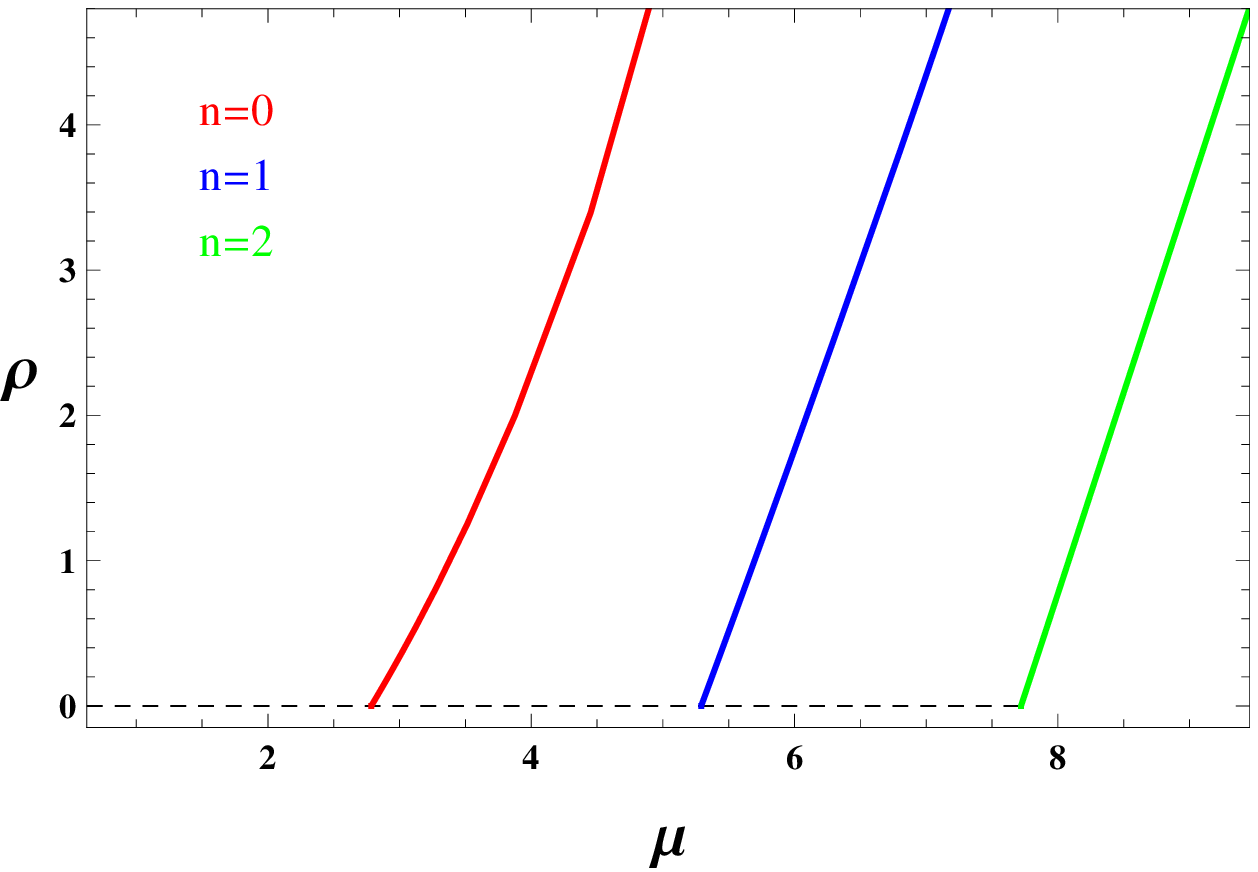}\\ \vspace{0.0cm}
\caption{\label{RealizationPWave} (Color online) The charge density $\rho$ as a function of the chemical potential $\mu$ with fixed masses of the vector field $m^{2}=0$ (left) and $m^{2}=5/4$ (right) when $\langle\mathcal{O}_{x}\rangle\neq0$. In each panel, the three lines from left to right correspond to the ground ($n=0$, red), first ($n=1$, blue) and second ($n=2$, green) excited states, respectively.}
\end{figure}

In Fig. \ref{RealizationPWave}, we plot the charge density $\rho$ as a function of the chemical potential with $m^{2}=0$ and $m^{2}=5/4$ from the ground state to the second excited state, which implies that for all the states, there is a critical chemical potential $\mu_{c}$ above which the system becomes unstable to develop vector hair leading to a second order phase transition in the dual field theory. By fitting these curves in the vicinity of the critical point, we observe that for the vector field mass $m^{2}=0$
\begin{eqnarray}\label{PWaveNRhoM0}
\rho\approx \left\{
                                      \begin{array}{ll}
                                        1.754\left(\mu-\mu^{(0)}_{c}\right), & \hbox{Ground state,} \\
                                        2.646\left(\mu-\mu^{(1)}_{c}\right), & \hbox{1st excited state,} \\
                                        2.896\left(\mu-\mu^{(2)}_{c}\right), & \hbox{2st excited state,}
                                      \end{array}
            \right.
\end{eqnarray}
and for the vector field mass $m^{2}=5/4$
\begin{eqnarray}\label{PWaveRhoNM125}
\rho\approx \left\{
                                      \begin{array}{ll}
                                        1.517\left(\mu-\mu^{(0)}_{c}\right), & \hbox{Ground state,} \\
                                        2.443\left(\mu-\mu^{(1)}_{c}\right), & \hbox{1st excited state,} \\
                                        2.790\left(\mu-\mu^{(2)}_{c}\right), & \hbox{2st excited state,}
                                      \end{array}
            \right.
\end{eqnarray}
where the critical chemical potentials $\mu^{(0)}_{c}$, $\mu^{(1)}_{c}$ and $\mu^{(2)}_{c}$ are given in Table \ref{PWave} for $m^{2}=0$ and $m^{2}=5/4$, corresponding to the ground, first and second excited states, respectively. Similar to the s-wave case, we again obtain the linear relation between the charge density and chemical potential near $\mu^{(n)}_{c}$ for the $n$-th excited state in the p-wave holographic insulator/superconductor model, i.e., $\rho\sim\left(\mu-\mu^{(n)}_{c}\right)$, which is independent of the vector field mass $m^{2}$ and the number of nodes $n$.

\subsection{Analytical investigation}

We have used the shooting method to numerically study the p-wave holographic insulator/superconductor phase transition with the excited states. Now we are in a position to investigate this p-wave holographic model by using the Sturm-Liouville method \cite{Siopsis,SiopsisBF} and analytically confirm the numerical findings.

\subsubsection{Critical chemical potential}

Changing the coordinate from $r$ to $z$ by $z=r_{s}/r$, we can express Eqs. (\ref{PWaveRhox}) and (\ref{PWaveAt}) in the $z$ coordinate as
\begin{eqnarray}\label{PWaveRhoxz}
\rho_{x}''+\left( \frac{1}{z}+\frac{f'}{f}\right)\rho_{x}'+\left[\frac{1}{z^{2}f}\left(\frac{qA_{t}}{r_{s}}\right)^{2}-\frac{m^{2}}{z^{4}f}\right]\rho_{x}=0,
\end{eqnarray}
\begin{eqnarray}\label{PWaveAtz}
A_{t}''+\left( \frac{1}{z}+\frac{f'}{f}\right)A_{t}'-\frac{2}{z^{2}f}\left(\frac{q\rho_{x}}{r_{s}}\right)^{2}A_{t}=0,
\end{eqnarray}
with the function $f(z)=(1-z^{4})/z^2$. Here and hereafter in this section the prime denotes the derivative with respect to $z$.

Considering the vector field $\rho_{x}=0$ at the critical chemical potential $\mu_{c}$ for the the ground and excited states, just as shown in Figs. \ref{CondPWave} and \ref{RealizationPWave}, we can obtain the physical solution $A_{t}(z)=\mu$ to Eq. (\ref{PWaveAtz}) when $\mu<\mu_{c}$, which takes the same form as that in the s-wave holographic insulator/superconductor model. Thus, assuming that $\rho_{x}$ takes the form
\begin{eqnarray}\label{PWaveRhoFz}
\rho_{x}(z)\simeq\frac{\langle{\cal O}_{x}\rangle}{r_{s}^{\Delta}}z^{\Delta}F(z),
\end{eqnarray}
with the boundary condition $F(0)=1$ satisfying the boundary behavior of $\rho_{x}$ in (\ref{PWaveinfinity}), we get the
equation of motion for the trial function $F(z)$
\begin{eqnarray}\label{PWaveFzmotion}
(QF')'+Q\left[P+V\left(\frac{q\mu}{r_{s}}\right)^{2}\right]F=0,
\end{eqnarray}
with
\begin{eqnarray}
Q=z^{1+2\Delta}f,~~~~~~P=\frac{\Delta}{z}\left(\frac{\Delta}{z}+\frac{f'}{f}\right)-\frac{m^{2}}{z^{4}f},
\end{eqnarray}
where $V(z)$ has been introduced in (\ref{TUV}). Following the standard procedure for the Sturm-Liouville eigenvalue problem \cite{Gelfand-Fomin}, we can deduce the eigenvalues of $q\mu/r_{s}$ from variation of the following function
\begin{eqnarray}\label{PWaveNewFzmotion}
\left(\frac{q\mu}{r_{s}}\right)^{2}=\frac{\int^{1}_{0}Q\left(F'^{2}-PF^{2}\right)dz}{\int^{1}_{0}QVF^{2}dz}.
\end{eqnarray}
Since $Q(1)\equiv0$ and $Q(0)\equiv0$ for the case of $\Delta=1+\sqrt{1+m^{2}}$ with the mass beyond the Breitenlohner-Freedman (BF) bound $m^{2}_{BF}=-1$ \cite{Breitenloher}, just as discussed in Sec. II for the s-wave holographic insulator/superconductor phase transition with the excited states, the condition $[Q(z)F(z)F'(z)]|_{0}^{1}=0$ can be satisfied automatically for the vector operator $\mathcal{O}_{x}$. This means that we shall require $F(z)$ to satisfy the Dirichlet boundary condition $F(0)=1$ rather than the Neumann boundary condition $F'(0)=0$. Thus, we expand the trial function $F(z)$ up to the eighth order, i.e., $F(z)=1-\sum_{k=1}^{k=8}a_{k}z^{k}$ for the operator $\mathcal{O}_{x}$ in the following calculation.

\begin{table}[ht]
\begin{center}
\caption{\label{PWaveOxM0}
The dimensionless critical chemical potential $q\mu_{c}/r_{s}$ obtained by the Sturm-Liouville method for the vector operator $\mathcal{O}_{x}$ and corresponding value of $a_{k}$ for the trial function $F(z)=1-\sum_{k=1}^{k=8}a_{k}z^{k}$ with the fixed mass of the vector field $m^{2}=0$ from the ground state to the sixth excited state in the p-wave holographic insulator and superconductor model.}
\begin{tabular}{c |c| c c c c c c c c c}
\hline
~$n$~&~$q\mu_{c}/r_{s}$~&~$a_{1}$~&~$a_{2}$~&~$a_{3}$~&~$a_{4}$~&~$a_{5}$~&~$a_{6}$~&~$a_{7}$~&~$a_{8}$~  \\
\hline
~$0$~ &~2.265~~&~0.000(9)~~&~0.631~~&~0.061~~&~-0.662~~&~~0.321~~&~0.180~~&~~-0.211~~&~0.056~ \\
\hline
~$1$~ &~4.741~~&~0.008~~&~2.715~~&~0.565~~&~-4.803~~&~3.243~~&~0.413~~&~~-1.204~~&~0.362~ \\
\hline
~$2$~ &~7.156~~&~0.066~~&~5.570~~&~5.006~~&~-30.887~~&~31.894~~&~-10.294~~&~~-1.908~~&~1.355~ \\
\hline
~$3$~ &~9.561~~&~0.424~~&~5.866~~&~34.438~~&~-165.470~~&~247.030~~&~-167.463~~&~~51.471~~&~-5.151~ \\
\hline
~$4$~ &~11.962~~&~1.633~~&~-5.082~~&~150.541~~&~-677.012~~&~1269.722~~&~-1185.247~~&~~546.634~~&~-100.302~ \\
\hline
~$5$~ &~14.363~~&~1.582~~&~-5.268~~&~271.599~~&~-1523.813~~&~3539.192~~&~-4099.436~~&~~2343.249~~&~-526.010~ \\
\hline
~$6$~ &~16.836~~&~-29.859~~&~400.377~~&~-1649.080~~&~2731.816~~&~-809.668~~&~-2966.267~~&~~3506.318~~&~-1182.820~ \\
\hline
\end{tabular}
\end{center}
\end{table}

\begin{table}[ht]
\begin{center}
\caption{\label{PWaveOxM125}
The dimensionless critical chemical potential $q\mu_{c}/r_{s}$ obtained by the Sturm-Liouville method for the vector operator $\mathcal{O}_{x}$ and corresponding value of $a_{k}$ for the trial function $F(z)=1-\sum_{k=1}^{k=8}a_{k}z^{k}$ with the fixed mass of the vector field $m^{2}=5/4$ from the ground state to the sixth excited state in the p-wave holographic insulator and superconductor model.}
\begin{tabular}{c |c| c c c c c c c c c}
\hline
~$n$~&~$q\mu_{c}/r_{s}$~&~$a_{1}$~&~$a_{2}$~&~$a_{3}$~&~$a_{4}$~&~$a_{5}$~&~$a_{6}$~&~$a_{7}$~&~$a_{8}$~  \\
\hline
~$0$~ &~2.785~~&~0.002~~&~0.751~~&~0.131~~&~-1.010~~&~~0.653~~&~0.042~~&~~-0.191~~&~0.057~ \\
\hline
~$1$~ &~5.291~~&~0.017~~&~2.614~~&~0.994~~&~-6.249~~&~5.280~~&~-1.092~~&~~-0.604~~&~0.258~ \\
\hline
~$2$~ &~7.720~~&~0.118~~&~4.651~~&~7.159~~&~-35.775~~&~41.522~~&~-20.472~~&~~3.455~~&~0.219~ \\
\hline
~$3$~ &~10.133~~&~0.614~~&~3.157~~&~40.026~~&~-170.718~~&~259.997~~&~-191.586~~&~~69.476~~&~-9.887~ \\
\hline
~$4$~ &~12.539~~&~1.799~~&~-7.103~~&~138.588~~&~-588.860~~&~1085.574~~&~-1015.018~~&~~475.456~~&~-89.491~ \\
\hline
~$5$~ &~14.944~~&~-0.456~~&~15.219~~&~140.116~~&~-1021.532~~&~2485.477~~&~-2908.675~~&~~1664.270~~&~-373.370~ \\
\hline
~$6$~ &~17.410~~&~-245.676~~&~2760.984~~&~-11762.316~~&~24627.132~~&~-26109.296~~&~11608.072~~&~~604.182~~&~-1482.480~ \\
\hline
\end{tabular}
\end{center}
\end{table}

In Tables \ref{PWaveOxM0} and \ref{PWaveOxM125}, we show the critical chemical potentials from the ground state to the sixth excited state by using the expression (\ref{PWaveNewFzmotion}) to compute the extremal values, which are in good agreement with the numerical results presented in Table \ref{PWave}. This suggests that the Sturm-Liouville method, by including higher orders of $z$ in the trial function $F(z)$, is very powerful to study the excited states of the p-wave holographic insulator/superconductor phase transition. Moreover, from these two tables, we observe that the critical chemical potential increases almost linearly when $n$ increases, which is consistent with the numerical finding given in Table \ref{PWave}. From the analytical results, we obtain
\begin{eqnarray}\label{PWaveMucAna}
\frac{q\mu_{c}}{r_{s}} \approx\left\{
                                      \begin{array}{ll}
                                        2.420n+2.295, &  \quad {\rm for} \  m^{2}=0\,,\\ \\
                                        2.429n+2.832, &  \quad {\rm for} \  m^{2}=5/4\,,
                                      \end{array}
                                    \right.
\end{eqnarray}
which agrees well with the numerical results shown in Eq. (\ref{PWaveMuc}). It is interesting to note that, from Eqs. (\ref{SWaveMuc}) and (\ref{SWaveMucAna}) for the s-wave holographic insulator/superconductor phase transition and Eqs. (\ref{PWaveMuc}) and (\ref{PWaveMucAna}) for the p-wave case, the difference of the dimensionless critical chemical potential $q\mu_{c}/r_{s}$ between the consecutive states is about 2.4, which is independent of the mass of the field and the type of the holographic model.

\begin{figure}[H]
\includegraphics[scale=0.65]{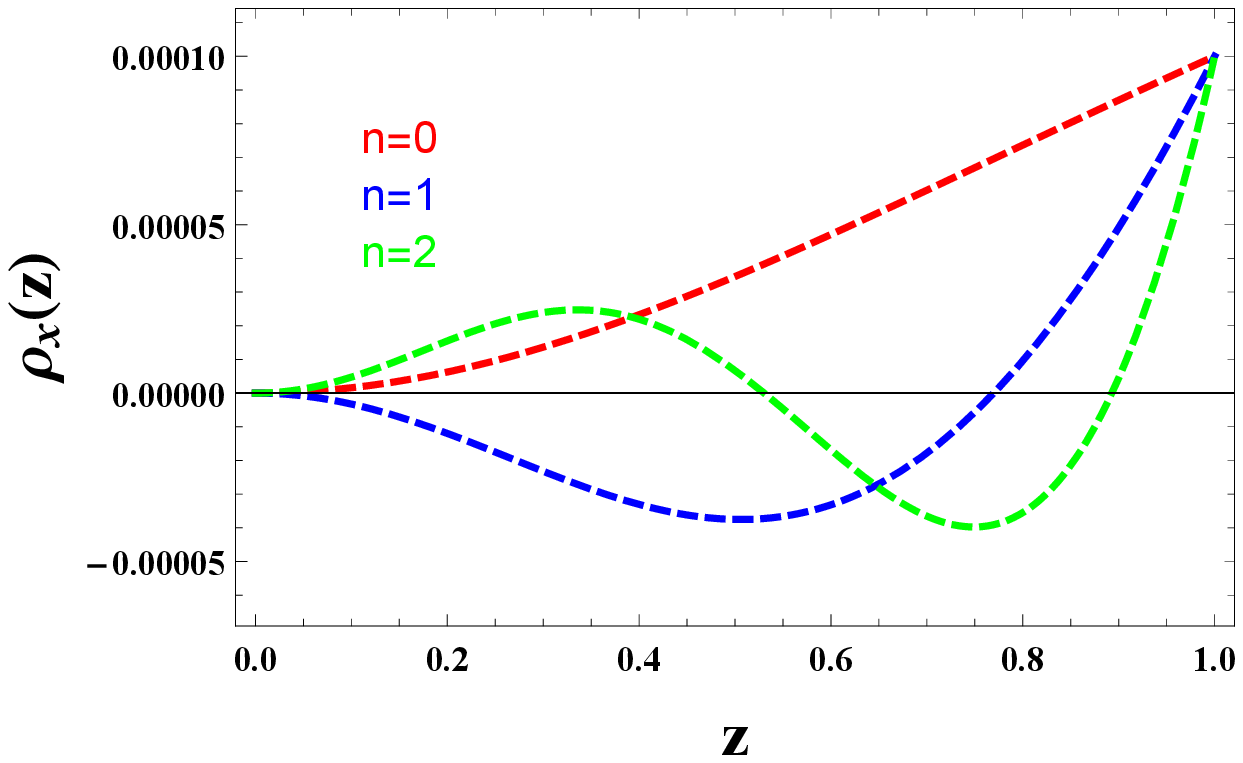}\hspace{0.2cm}%
\includegraphics[scale=0.65]{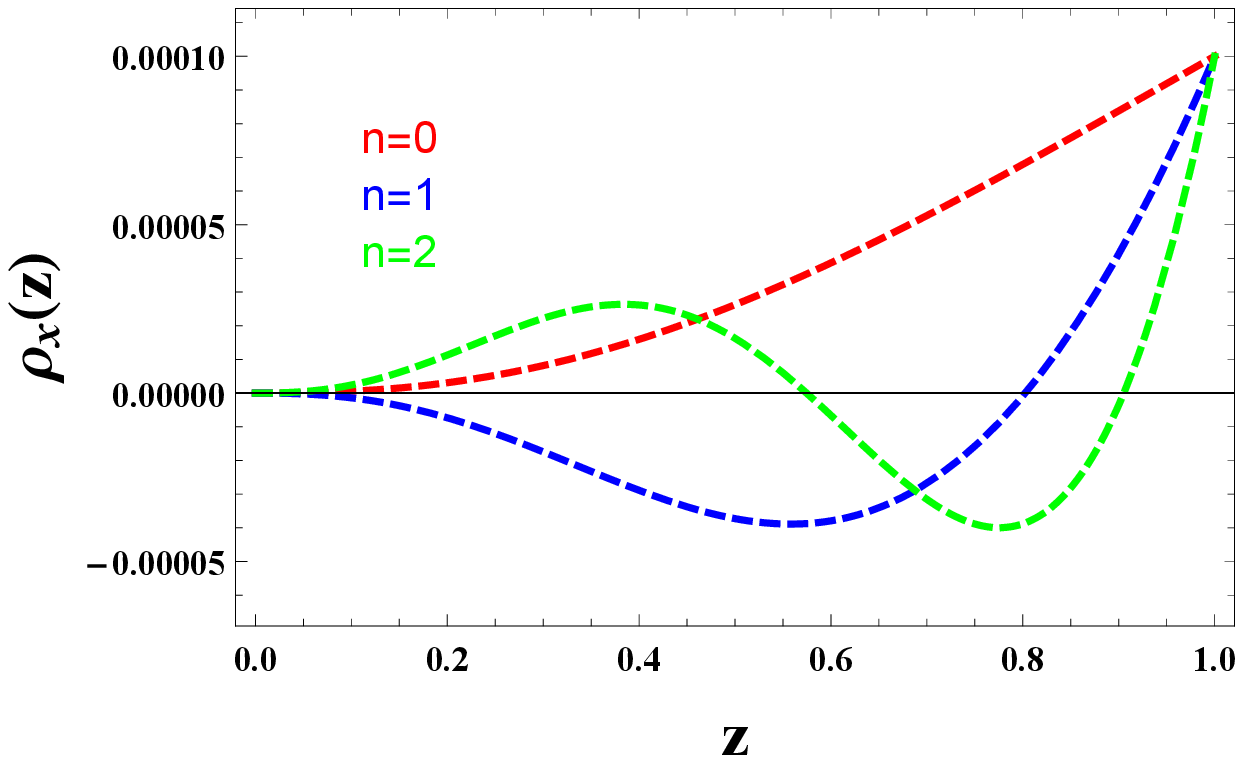}\\ \vspace{0.0cm}
\caption{\label{PWRhoxzAna} (Color online) The vector field $\rho_{x}(z)$ as a function of the radial coordinate $z$ outside the horizon with the vector operator $\mathcal{O}_{x}$ for the fixed masses of the vector field $m^{2}=0$ (left) and $m^{2}=5/4$ (right) by using the analytical Sturm-Liouville method with $\rho_{x}(1)=0.0001$ in Eq. (\ref{PWaveRhoFz}). In each panel, the three lines from top to bottom correspond to the ground ($n=0$, red), first ($n=1$, blue) and second ($n=2$, green) excited states, respectively.}
\end{figure}

Also, we can use the expression (\ref{PWaveRhoFz}) to plot the distribution of the vector field $\rho_{x}(z)$ as a function of $z$ for the vector operator $\mathcal{O}_{x}$ with the fixed masses of the vector field $m^{2}=0$ and $m^{2}=5/4$ by setting the initial condition $\rho_{x}(1)=0.0001$, just as shown in Fig. \ref{PWRhoxzAna}. Compared with the numerical results shown in Fig. \ref{PWRhoxz}, the agreement of the analytical results derived from the Sturm-Liouville method with the numerical calculation is impressive. Thus, we conclude that the improved Sturm-Liouville method can not only analytically calculate the critical chemical potential of the excited states, but also study the behaviors of the vector field near the critical point.

\subsubsection{Critical phenomena}

In the vicinity of the critical point, we may expand $A_{t}(z)$ in small $\langle{\cal O}_{x}\rangle$ as
\begin{eqnarray}
A_{t}(z)\sim\mu_{c}+2\mu_{c}\left(\frac{q\langle{\cal O}_{x}\rangle}{r_{s}^{1+\Delta}}\right)^{2}\chi(z)+\cdots,
\end{eqnarray}
with the boundary condition $\chi(1)=0$ at the tip. Then, with the help of Eqs. (\ref{PWaveAtz}) and (\ref{PWaveRhoFz}),
we obtain
\begin{eqnarray}\label{PWaveOx}
(W\chi')'-z^{2\Delta-1}F^{2}=0,
\end{eqnarray}
where $W(z)=zf(z)$ has been defined in the s-wave holographic insulator/superconductor model.

Following the strategy utilized for the analysis regarding the critical phenomena in the s-wave holographic insulator/superconductor phase transition, near $z\rightarrow0 $ we can expand $A_{t}$ as
\begin{eqnarray}\label{PWaveexpandingAt}
A_{t}(z)\simeq\mu-\frac{\rho }{r_{s}^{2}}z^{2}\simeq\mu_{c}+2\mu_{c}\left(\frac{q\langle{\cal
O}_{x}\rangle}{r_{s}^{1+\Delta}}\right)^{2}\left[\chi(0)+\chi'(0)z+\frac{1}{2}\chi''(0)z^{2}+\cdots\right].
\end{eqnarray}
From the coefficients of the $z^{0}$ term in both sides of the above formula, we get
\begin{eqnarray}\label{PWaveOxSL}
\frac{q\langle{\cal
O}_{x}\rangle}{r_{s}^{1+\Delta}}=\frac{1}{\sqrt{2\chi(0)}}\left(\frac{\mu}{\mu_{c}}-1\right)^{1/2},
\end{eqnarray}
where $\chi(0)=c_{4}-\int_{0}^{1}W^{-1}[c_{5}+\int_{1}^{z}x^{2\Delta-1}F(x)^{2}dx]dz$ with the integration constants $c_{4}$ and $c_{5}$ which can be determined by the boundary condition of $\chi(z)$ in Eq. (\ref{PWaveOx}). As an example, for the vector field mass $m^{2}=0$ we observe that
\begin{eqnarray}\label{PWaveOxM0SL}
\frac{q\langle{\cal O}_{x}\rangle}{r_{s}^{3}}\approx \left\{
                                      \begin{array}{ll}
                                        2.938\left(\frac{\mu}{\mu^{(0)}_{c}}-1\right)^{1/2}, & \hbox{Ground state,} \\
                                        8.981\left(\frac{\mu}{\mu^{(1)}_{c}}-1\right)^{1/2}, & \hbox{1st excited state,} \\
                                        16.717\left(\frac{\mu}{\mu^{(2)}_{c}}-1\right)^{1/2}, & \hbox{2st excited state,}
                                      \end{array}
                                    \right.
\end{eqnarray}
where the critical chemical potentials $\mu^{(0)}_{c}$, $\mu^{(1)}_{c}$ and $\mu^{(2)}_{c}$ are given in Table \ref{PWaveOxM0}, which correspond to the ground, first and second excited states, respectively. Obviously, Eq. (\ref{PWaveOxM0SL}) can be compared with the numerical results given in Eq. (\ref{PWaveOxNM0}). Since Eq. (\ref{PWaveOxSL}) is valid in general, near the critical point we obtain $\langle{\cal O}_{x}\rangle\sim\left(\mu-\mu^{(n)}_{c}\right)^{1/2}$, which analytically confirms that, for all the excited states, the phase transition between the p-wave insulator and superconductor is of the second order and the critical exponent of the system attains the mean-field value $1/2$.

According to the coefficients of the $z^{2}$ term in Eq. (\ref{PWaveexpandingAt}), we finally have
\begin{eqnarray}\label{PWaveRhoAna}
\frac{\rho}{r_{s}^{2}}=-\mu_{c}\chi''(0)\left(\frac{q\langle{\cal
O}_{x}\rangle}{r_{s}^{1+\Delta}}\right)^{2}=\Gamma(m,n)(\mu-\mu_{c}),
\end{eqnarray}
where $\Gamma(m,n)=[2\chi(0)]^{-1}\int_{0}^{1}z^{2\Delta-1}F^{2}dz$ is a function of the vector field mass $m^{2}$ and the number of nodes $n$. For the case of $m^{2}=0$, we get
\begin{eqnarray}
\frac{\rho}{r_{s}^{2}}\approx \left\{
                                      \begin{array}{ll}
                                        1.128\left(\mu-\mu^{(0)}_{c}\right), & \hbox{Ground state,} \\
                                        1.247\left(\mu-\mu^{(1)}_{c}\right), & \hbox{1st excited state,} \\
                                        1.262\left(\mu-\mu^{(2)}_{c}\right), & \hbox{2st excited state,}
                                      \end{array}
                                    \right.
\end{eqnarray}
which is again consistent with the numerical result shown in Eq. (\ref{PWaveNRhoM0}). Here the critical chemical potentials $\mu^{(0)}_{c}$, $\mu^{(1)}_{c}$ and $\mu^{(2)}_{c}$ have been presented in Table \ref{PWaveOxM0} for the ground, first and second excited states, respectively. Therefore, from Eq. (\ref{SWaveRhoAna}) for the s-wave holographic insulator/superconductor phase transition and Eq. (\ref{PWaveRhoAna}) for the p-wave case, we analytically confirm that a linear relationship $\rho\sim\left(\mu-\mu^{(n)}_{c}\right)$ exists between the charge density and chemical potential near $\mu^{(n)}_{c}$ for the $n$-th excited state, which agrees well with the numerical calculation for both the s-wave and p-wave holographic insulator/superconductor phase transitions with the excited states.

\section{conclusions}

In this work, we have presented a family of solutions of the holographic insulator/superconductor phase transitions with the excited states in the probe limit, by using the numerical shooting method and analytical Sturm-Liouville method. In particular, we showed that the analytical results, obtained by including more higher order terms in the expansion of the trial function in the Sturm-Liouville method, are in good agreement with the numerical data. Interestingly, we noticed that this improved analytical method can not only analytically investigate the holographic insulator/superconductor phase transitions with the excited states but also the behaviors of the condensed fields near the critical point of the phase transition, which implies that the Sturm-Liouville method is a robust method to disclose the properties of the insulator/superconductor phase transition systems even for the excited states.  For both the s-wave (scalar field) and p-wave (vector field) insulator/superconductor models, we found that the critical chemical potential increases linearly as the number of nodes increases, which means that the excited state has a higher critical chemical potential than the corresponding ground state. It should be noted that, although the underlying mechanism is still unclear, the difference of the dimensionless critical chemical potential between the consecutive states is around 2.4 regardless of the type of the holographic model, which is obviously different from the finding of the metal/superconductor phase transition in the backgrounds of AdS black hole where the difference is around 5.2 \cite{WangJHEP2020,QiaoEHS}. Moreover, for all the excited states in both s-wave and p-wave models, we observed that the phase transition of the systems belongs to the second order with the mean-field critical exponent $1/2$, and the charge density scales linearly with the chemical potential in the vicinity of the critical point. Since the backreaction can provide richer physics in the holographic insulator/superconductor models, it would be of great interest to generalize our study to the case where the backreaction is taken into account. We will leave it for further study.

\begin{acknowledgments}

We thank Professor Yong-Qiang Wang for his helpful discussions and suggestions. This work was supported by the National Natural Science Foundation of China under Grant Nos. 11775076, 11875025, 11705054, 12035005 and 11690034; Hunan Provincial Natural Science Foundation of China under Grant Nos. 2018JJ3326 and 2016JJ1012.

\end{acknowledgments}

\end{document}